\documentclass[useAMS,usenatbib,fleqn]{mn2e}
\usepackage{graphicx}
\usepackage{amsmath, amssymb, amsfonts}
\usepackage{color}
\usepackage{times}
\usepackage{enumerate} 
\usepackage{colordvi} 
\usepackage{bm}

\newcommand{\be}{\begin{eqnarray}}
\newcommand{\ee}{\end{eqnarray}}
\newcommand{\simgt}{\lower.5ex\hbox{$\; \buildrel > \over \sim \;$}}
\newcommand{\simlt}{\lower.5ex\hbox{$\; \buildrel < \over \sim \;$}}

\newcommand{\bfk}{{\bf k}}

\newcommand{\bfq}{{\bf q}}

\newcommand{\Omegam}{\Omega_{\rm m}}

\title[Anisotropic clustering of luminous red galaxies]{Simulating the anisotropic clustering of luminous red galaxies with subhalos: a direct confrontation with 
observation and cosmological implications}
\author[T. Nishimichi and A. Oka]
{Takahiro Nishimichi$^{1}$\thanks{E-mail:nishimic@iap.fr}
and Akira Oka$^{2}$
\\
$^1$Institut d'Astrophysique de Paris, 98 bis boulevard Arago, 75014 Paris, France\\
$^2$Department of Physics, School of Science, The University of Tokyo, Bunkyo-ku, Tokyo 113-0033, Japan\\
}

\begin{document}

\date{Accepted --. Received --; in original form --}

\pagerange{\pageref{firstpage}--\pageref{lastpage}} \pubyear{20??}

\maketitle
\label{firstpage}

%
\begin{abstract}
We model the apparent clustering anisotropy of Luminous Red Galaxies (LRGs) in the Sloan Digital Sky Survey 
using subhalos identified in cosmological $N$-body simulations.
We first conduct a Markov-chain Monte Carlo analysis on the parameters characterizing subhalos 
hosting LRGs assuming a specific $\Lambda$CDM cosmology on which we run the simulations.
We show that simple models with central and satellite subhalos can explain the observed
multipole moments of the power spectrum up to hexadecapole on large scales ($k\lesssim0.3\,h\mathrm{Mpc}^{-1}$). 
A satellite fraction of $20$ to $30$ per cent is favored weakly depending on the detail of the model. 
The fraction is shown to be robust when we adopt a more refined model based on the halo occupation number
from the literature.
We then vary cosmological parameters controlling the anisotropy in redshift-space effectively
by deforming the simulation box (the Alcock-Paczynski effect) and changing the amplitude of the
velocities (the redshift-space distortions). 
We demonstrate that we can constrain the geometry of the universe, the structure growth rate, 
and the parameters characterizing LRGs simultaneously.
This is a step toward cosmological analysis with realistic bias description
beyond empirical bias functions with nuisance parameters.
\end{abstract}


\section{Introduction}
Understanding the nature of a population of galaxies is the key to derive unbiased
cosmological constraints using their spatial clustering pattern. This issue of galaxy bias is one of 
the biggest obstacles in the modern cosmology \citep{Kaiser84}. In other words, selecting a controlled 
sample of galaxies whose environmental properties are well understood is an appropriate
way for this purpose. Luminous Red Galaxies (LRGs) collected by the
Sloan Digital Sky Survey (SDSS, \citealt{SDSS}) are believed to be such a galaxy sample 
\citep{Eisenstein01}: a population mostly composed of central galaxies associated with massive halos.
This understanding can be inferred from a number of observational facts such as 
a low number density, a large fraction of single-LRG 
systems \citep{Reid09}, and a high bias factor at large scale relative 
to the clustering amplitude of the underlying matter distribution \citep{Eisenstein05}. 
All these observed features make them useful for cosmological
applications, and indeed a lot of important cosmological implications have been derived
using this sample including the first clear detection of baryon acoustic oscillations (BAOs) by \citet{Eisenstein05}.

One of the most popular analytical approaches to model the clustering of galaxies is so-called the 
halo occupation distribution (HOD) approach based on the halo model 
(e.g., \citealt{Ma:2000lr,Peacock:2000qy,Seljak:2000uq,Scoccimarro:2001fj,Berlind:2002kx}). 
One assumes that all the observed galaxies 
live in halos, and one can calculate the clustering properties once the probability of having $N$ galaxies in a halo with mass 
$M_\mathrm{host}$ is given. One usually assumes a simple functional form for the (mean) halo occupation number as
a function of the halo mass, and determines the model parameters by fitting to some observed properties. 
In particular, the HOD parameters have been investigated for LRGs based on the spatial clustering on relatively small scales
\citep{Kulkarni07, White07, Blake08, Wake08, Padmanabhan09a, Zheng09}, galaxy-galaxy lensing
\citep{Mandelbaum06,Hikage13a},
or a direct measurement of the number distribution of LRGs forming groups \citep{Ho09,Reid09}. 
LRGs may also be able to be modeled with simulated subhalos employing
abundance matching schemes (e.g., \citealt{Conroy06,Masaki13}).

On another front, there have been a lot of attempts to model the statistics of galaxy spatial clustering at the scale of BAOs
based on perturbation theory (PT; see \citealt{Bernardeau02} for a thorough review of the standard PT 
and also \citealt{Jeong06,Nishimichi07,Eisenstein07,Jeong09b,Padmanabhan09b,Sherwin12,McCullagh13} for its application 
to BAOs) beyond linear theory in the light of ongoing/near future galaxy redshift surveys.
Since these survey projects aim at a precise determination of cosmological distances, the required accuracy
in the theoretical modeling is highly demanding. In such a situation, a number of ``renormalized" PT
techniques have been developed to have a better convergence of perturbative series
expansions on statistical quantities at the scale of interest 
\citep{Crocce06b,Crocce06c,Matarrese07,McDonald07,Valageas07,Crocce08,Taruya08,Matsubara08a,Matsubara08b,
Bernardeau08,Matarrese08a,Pietroni08,Hiramatsu09,Anselmi11,Okamura11,Sugiyama12a,Sugiyama12b,
Sugiyama13,Valageas13a,Bernardeau13}.
It is also worth mentioning the importance of cosmological $N$-body simulations in developing these analytical models.
Since all the perturbative schemes involve some approximation, ansatz or truncation, and moreover,
there are fundamental limitations of perturbative approaches such as the breakdown of the single-stream approximation 
after shell crossing \citep{Valageas11c,Valageas2013}, one has to confirm the validity of one's scheme and determine with 
care the applicable range in wavenumber and redshift by testing with fully nonlinear predictions based on simulations 
\citep{Crocce08,Nishimichi09,Taruya09,Carlson09,Valageas11a,Sato11}.
Recently, some codes to evaluate the nonlinear power spectrum of the cosmic density field very rapidly
based on new perturbative approaches have been made publicly available \citep{Crocce12,Taruya12}. 
These tools are practically useful in confronting the analytical predictions with observational data.

Despite all the recent progress, 
galaxy bias is still difficult to implement into these new techniques in a consistent manner
without loosing the non-perturbative properties in those theories (some recent attempts along this line
can be found in e.g., \citealt{McDonald06,Matsubara08b,Nishizawa13}). 
Also, the effect of one-halo term (or, equivalently, satellite galaxies) could be significant on 
the power spectrum especially in redshift space as recently suggested by \citet{Hikage13b}, which
is beyond the scope of PT calculations.
Thus it is not straightforward to analyze {\it galaxy} clustering with renormalized PT techniques and extract
cosmological information robustly, even if the environmental properties of the observed galaxy sample,
such as those of LRGs mentioned above, are very well understood.

The purpose of this study is to see weather the current state-of-the-art $N$-body simulations can explain the 
clustering of galaxies (LRGs, more specifically) in redshift space on large scales where most of the cosmological
information exists. Also, we explore the possibility of extracting cosmological information by confronting 
simulations with observation instead of using an analytical model. Studies along this line is not straightforward
for various reasons. One is from the fact that we simulate {\it one realization} of the cosmological random field in finite
volume drawn from an assumed cosmological model while we observes {\it another realization} under 
the correct cosmology.
Another is the high computational cost to cover the multi-dimensional cosmological parameter space with simulations.
Since this parameter space has typically as many as six dimensions in the standard $\Lambda$CDM model 
(and even more when one wishes to test some non-standard models),
this is not realistic with high-resolution simulations with sufficient volume.
Studies on the rescaling of the simulation outputs to different cosmological models can be found in e.g.,
\citet{Tormen96,Cole97,Angulo10,Mead13}, and its applications to extract cosmological information or to infer
the cosmological model dependence from galaxy properties are performed in \citet{Simha13,Guo13}.

We partly overcome these difficulties in this study by employing the following methodology.
First, by taking an ensemble average over different random realizations of simulations whose total volume
is much larger than the observed volume, we obtain a well converged prediction that can directly be
compared with observation as one does with analytical models.
We next introduce three parameters, which affect the apparent {\it anisotropy} of the clustering and are closely related
to some cosmological parameters, and float them to see their impact on the power spectrum without re-running new simulations 
starting from new initial conditions.

Our new parameters are responsible for the growth rate of the cosmic perturbations, 
the Hubble parameter and the angular diameter distance at the effective redshift of the observed galaxies.
The first one can be observed through the redshift-space distortions (RSDs) caused by peculiar velocities of galaxies
\citep{Kaiser87}.
Our first parameter, that scales the amplitude of velocities in simulation outputs, 
amplifies or suppresses the magnitude of RSDs. 
Since the significance of the anisotropy induced by RSDs is a good indicator of the growth rate of the cosmic structure,
$f\sigma_8$, where $f\equiv \mathrm{d}\ln D_+/\mathrm{d}\ln a$ with $D_+$ being the linear growth factor,
we can test the underlying gravity theory by measuring it (e.g., 
\citealt{Percival04,Linder08,Guzzo08,Yamamoto08,Song09,Blake11a,Reid12,Beutler12,Samushia13}).
The other two parameters deform the simulation box and induce apparent anisotropy to the clustering of the mock galaxies.
By doing this, we simulate the Alcock-Paczynski effect (\citealt{Alcock79}; the AP effect, hereafter), which should be there
in the observed clustering if the cosmological model assumed in the conversion of redshifts to 
the three-dimensional positions has a mismatch with the true one.
One can determine the Hubble parameter, $H$, and the angular diameter distance, $D_\mathrm{A}$,
through this effect combined with the characteristic scale of BAOs 
(e.g., \citealt{Matsubara:1996qy,Ballinger96,Hu03,Okumura08,Padmanabhan08,Shoji09,Taruya11,Blake11c,Blake12,Reid12,Chuang12,Xu13,Anderson:2014kq,Kazin13,Sanchez13,Chuang13a,Chuang13b}).
Although more involved approaches such as one in \citet{Angulo10} may rescale the simulations more 
accurately, our simple method is computationally very easy to implement and 
can be safely applied as far as one focuses on the anisotropy of the clustering.

An important question that we would like to ask here is whether we can distinguish these distortions
with uncertainties in the modeling of galaxies, in particular velocities of galaxies.
Adopting different prescriptions for the LRG-subhalo connection, we show how much the resultant cosmological 
constraints are affected.
In the accompanying paper \citep{Oka13}, we present a similar analysis using the same observed dataset but with 
an analytical model for the nonlinear galaxy power spectrum in redshift space.
In the model, the galaxy bias as well as the Fingers-of-God suppression of the power spectrum \citep{Jackson72} 
are modeled by rather simple functional forms with free parameters without much physical justification at the quantitative level.
On the other hand, we expect that our model based on simulated subhalos physically describes these effects 
in a fully nonlinear manner as long as the gravitational interaction is concerned.
Given these differences, it would also be interesting to discuss the consistency between the cosmological 
parameters derived with two fundamentally different prescriptions for LRGs.

This paper is organized as follows.
We first briefly explain the observed power spectrum in Section~\ref{sec:data}.
We then discuss our methods to model LRGs using simulated subhalos in Section~\ref{sec:sim}.
Our main results are presented in Section~\ref{sec:results}. We compare our cosmological constraints with those in
\citet{Oka13} to check the consistency of the two analyses in that section.  A comparison with a model based on
the observed HOD is also discussed.
We summarize the findings of this study and add some discussion on the possible future generalization of our method
toward a fully consistent cosmological parameter estimation in Section~\ref{sec:discussion}.

\section{Data}
\label{sec:data}
In this section, we briefly explain the galaxy sample analyzed in this study and the measurement of the power spectrum
done in \citet{Yamamoto10}.
The LRGs used in the measurement is drawn from the Seventh Data Release of 
the Sloan Digital Sky Survey \citep{SDSSDR7}. $100,157$ LRGs with spectra in the redshift range of $0.16<z<0.47$
are selected from the Northern Galactic Cap covering $\sim 7,150\,\mathrm{deg}^2$.
The anisotropic power spectrum is expanded into multipole moments, 
\be
P(k,\mu_\mathbf{k}) = \sum_{\ell=\mathrm{even}}\mathcal{P}_\ell(\mu_\mathbf{k})P_\ell(k),
\ee
where $\mu_\mathbf{k}$ is the directional cosine
of the wavevector with respect to the line of sight and $\mathcal{P}_\ell(\mu_\mathbf{k})$ denotes the Legendre polynomial.
The multipole moments of the power spectrum, $P_\ell(k)$, are measured with the procedure developed in \citet{Yamamoto06}
at the effective redshift of $z=0.3$.  
We analyze the moments up to hexadecapole ($\ell=4$) on large scales ($k\leq0.305\,h\,\mathrm{Mpc}^{-1}$).
The estimation of the statistical error on the measured spectrum is also described in \citet{Yamamoto06}, 
in which they employ a method based on \citet{Feldman94} assuming Gaussianity of the density field. 
Although different multipole moments have non-zero cross covariance 
even when the underlying density field obeys the Gaussian statistics, we ignore it in this study for simplicity.
At leading order, this contribution is proportional to $\beta=f/b$, where $b$ is the linear bias factor, 
and is less important for highly biased tracers \citep{Taruya11}. Since the LRGs used in this analysis 
have bias as large as $\sim 2$, ignoring the cross covariance would not affect the final result significantly.

Although we basically analyze the power spectrum measured from the same dataset as in our accompanying paper, 
\citet{Oka13}, we employ a different assumption in the underlying cosmology when redshifts of galaxies are 
converted to distances.
In this study, we adopt a flat $\Lambda$CDM cosmology with $\Omegam = 0.28$ and $h=0.7$ that is exactly the
same as in \citet{Yamamoto10}, while \citet{Oka13} re-measure the
power spectrum with $\Omegam = 0.32$ and $h=0.67$ motivated by the recent PLANCK result \citep{Planck_cosmo}.
Since the cosmological model used in the simulations are closer to the former, we
simply adopt the original measurement by \citet{Yamamoto10}.
See Table~\ref{tab:cosmo} for cosmological models discussed in this study.
In Section~\ref{sec:results}, we discuss the consistency between the derived cosmological parameters and 
those assumed here (and in the $N$-body simulations) and propose a possible iterative scheme to perform
a fully consistent cosmological analysis in Section~\ref{sec:discussion}.

\begin{table*}
\begin{minipage}{135mm}
\begin{center}
\caption{Summary of the cosmological parameters of flat $\Lambda$CDM model.
The first row shows the model assumed for the distance-redshift relation in the measurement of the power spectrum.
The parameters in the second row are adopted in the $N$-body simulations.
The third to the sixth are the maximum likelihood parameters from CMB observations without combining with
other experiments, and they are 
plotted in Figs~\ref{fig:cosmo_raw} and \ref{fig:cosmo_raw2} to see the consistency with our analysis.
\label{tab:cosmo}}
\begin{tabular}{l c c c c c c c c}
\hline 
 & $\Omegam$ & $h$ & $\sigma_8$ & $r_\mathrm{s}^a$ & $H^b$ 
 & $D_\mathrm{A}^c$ & $f\sigma_8^d$ & reference\\
\hline 
dist.-$z$ rel. & 0.28 & 0.7 & - & - & 80.88 & 922.81 & - &\citet{Yamamoto10} \\
sim. & 0.279 & 0.701 & 0.817 & 104.02 & 80.88 & 921.69 & 0.462 & \citet{Nishimichi11} \\ 
WMAP5 & 0.249 & 0.725 & 0.787 & 108.74 & 82.57 & 897.51 & 0.420 & \citet{WMAP5}\\
WMAP7 & 0.271 & 0.703 & 0.801 & 105.08 & 80.88 & 920.91 & 0.448 & \citet{WMAP7}\\
WMAP9 & 0.282 & 0.697 & 0.820 & 103.46 & 80.59 & 926.46 & 0.461 & \citet{WMAP9}\\
PLANCK & 0.318 & 0.671 & 0.834 & 98.87 & 78.84 & 954.68 & 0.489 & \citet{Planck_cosmo}\\
\hline 
\end{tabular}

$^a$ The sound horizon scale at the baryon drag epoch in $h^{-1}$ Mpc.\\
$^b$ The Hubble parameter in $\mathrm{km/s/Mpc}$ evaluated at $z=0.3$.\\
$^c$ The angular diameter distance in Mpc at $z=0.3$.\\
$^d$ The growth-rate parameter, $f$, multiplied by $\sigma_8$. The value is computed at the redshift of the simulation 
output, $z=0.35$, for the model adopted in the simulations (second row), while this parameter is evaluated at $z=0.3$, 
the effective redshift of the survey, for the other models.
\end{center}
\end{minipage}
\end{table*}

\section{Modeling LRGs with simulations}
\label{sec:sim}
Here we describe our model of mock LRGs constructed from cosmological $N$-body simulations.
After showing the detail of the simulations and subhalos in Section~\ref{subsec:settings}, 
we explain how we connect them and mock LRGs in Section~\ref{subsec:model}.
We then summarize the method to measure the model power spectrum and to fit to the
observed data in Section~\ref{subsec:fit}. Our method to simulate the cosmological dependence of 
the apparent anisotropy is described in Section~\ref{subsec:RSD}.
\subsection{Simulations and subhalo identification}
\label{subsec:settings}
The cosmological $N$-body simulations used in this study are performed in \citet{Nishimichi11}.
Employing $1,280^3$ collisionless particles in periodic cubes with the side length of 
$1144.72\,h^{-1}$ Mpc, we simulate the gravitational growth of structure with a publicly-available
tree-PM code, \textsc{GADGET2} \citep{GADGET2}. The initial conditions are set by a code developed in 
\citet{Nishimichi09} and parallelized in \citet{Valageas11a} based on the second-order Lagrangian
perturbation theory (2LPT, e.g., \citealt{Scoccimarro98,Crocce06a}).
We assume a flat $\Lambda$CDM universe with the parameters derived by the
five-year observation by WMAP satellite (\citealt{WMAP5}; WMAP5+BAO+SNALL in the reference, 
and see also Table~\ref{tab:cosmo}) to compute the linear power spectrum using
\textsc{CAMB} \citealt{CAMB}. Fifteen independent random realizations are simulated and
snapshots at $z=0.35$ are stored.

Unfortunately, the spatial distribution of dark matter particles are available only for $11$ realizations
out of $15$ due to a problem in our hard disk. Using the remaining $11$ realizations, we identify subhalos
using an independent implementation of \textsc{SUBFIND} algorithm \citep{SUBFIND}. 
In the code, we first find Friends-of-Friends (FoF; e.g., \citealt{Davis:1985fk}) groups and then search for 
gravitationally bound particles inside each FoF group.
In this paper, we conventionally refer to the most massive subhalo in a FoF group as a {\it central}, while
the rest of the subhalos are called as {\it satellites}. We keep all the subhalos with mass larger than
$5.5\times10^{11}\,h^{-1}M_\odot$ ($10$ $N$-body particles) hosted by halos larger than 
$1.8\times10^{12}\,h^{-1}M_\odot$. These subhalos are used to reproduce the anisotropic
clustering of observed LRGs in what follows.

\subsection{Connecting subhalos to LRGs}
\label{subsec:model}
Parameters describing the properties of galaxies such as the HOD are often discussed using 
clustering measures on relatively small scales or one-point statistics. They are discussed
usually based on a specific cosmological model.
On the other hand, the large-scale clustering, often aiming at extraction of cosmological information, is expected
to be insensitive to the detail of the nature of galaxies. 
For example, a scale-independent linear bias model might be fine, though not fully validated, at the large scale limit. 
Since we here discuss the clustering in linear to weakly nonlinear regime, we wish to avoid introducing 
many parameters and/or employing a specific functional form for bias such as those for the HOD.
The determination of the detailed model parameters must be difficult unless we extend the analysis to 
sufficiently smaller scales.
We employ rather simple models based on LRG-(sub)halo connection and discuss the validity with the goodness-of-the-fit
to the observed data.

In this study, we examine several different descriptions of mock LRGs using subhalos identified above.
In each of the five models below, we have two options: the model ``a" assigns the center-of-mass position 
and velocity of a subhalo to a mock LRG, while in the model ``b" we regard 
the most bound $N$-body particle in a subhalo as a mock LRG. While the mock LRGs in these two models have 
almost the same position, the velocity can significantly be different. The modeling
of velocities of mock LRGs is crucially important in analyzing the anisotropic clustering because it directly affects the
signal of RSDs, and the two models investigated here serve as the two extreme cases; 
the motion of LRGs perfectly coincides with that of their host subhalos in model ``a", 
while the model ``b" takes account of the relative motion of mock LRGs with respect to their hosts. 
Naively, we expect that the velocity of a LRG aligns to that of the host subhalo.
However, assigning the center-of-mass velocities to mock LRGs (i.e., model a) is already a big assumption.
We thus supplementally infer the results of model b to see the impact of an unexpected velocity component, if exists, 
especially on the cosmological constraints.
In what follows we employ notations such as ``Model 3b" to label the total of ten models.

The five models we test in this study are as follows.
\begin{description}
\item{\bf Model 1: centrals only}\mbox{}\\
The first model assumes one-to-one correspondence between a mock LRG and a FoF group in the simulations.
Mock LRGs populate only in the most massive subhalos (i.e., centrals) in the host FoF groups in this model.
The only model parameter is the minimum host halo mass, $M_{\rm host, min}$.
This model is motivated by the observational fact that about $95$ per cent of LRGs do not have close 
companions \citep{Reid09}.
Though this model might be too naive, it is still interesting to see how well we can
reproduce the observed anisotropy of LRGs with the central population alone.

\item{\bf Model 2: centrals $+$ satellites}\mbox{}\\
A natural extension of Model 1 is to add satellites on top of centrals.
In addition to the parameter $M_\mathrm{host,min}$ in Model 1, this model has the second parameter, 
$M_{\rm sub, min}$, the minimum mass of subhalos needed to host mock LRGs.
The former mainly controls the environment of mock LRGs and thus the strength of bias, while the latter
effectively determines the fraction of satellites. Model 2 is the baseline model in this study.

\item{\bf Model 3: different criteria for centrals and satellites (minimum mass)}\mbox{}\\
The next model we consider treats centrals and satellites differently. Since we expect that
these two species of subhalos have different velocity structure, we expect that the fraction of satellites 
might be important to explain the observed anisotropic power spectrum in redshift space. In order to
test the robustness of the satellite fraction derived with Model 2, we introduce different minimum 
subhalo masses for the two species:
$M_{\rm cen, min}$ for centrals and $M_{\rm sat, min}$ for satellites. This model
has three parameters in total ($M_\mathrm{host, min}$, $M_{\rm cen, min}$ and $M_{\rm sat, min}$).

\item{\bf Model 4: different criteria for centrals or satellites (random sampling)}\mbox{}\\
As in Model 3, the fourth model also assigns mock LRGs to centrals and satellites in a different manner.
This time, we assign mock LRGs only to subhalos randomly chosen out of those that satisfy
the condition set by $M_\mathrm{host, min}$ and $M_\mathrm{sub, min}$.
Although we can in principle conduct a random sampling to both of centrals and satellites with different probabilities, 
we do so only either population at one time and keep all the subhalos of the other population to reduce the shot noise. 
This is because the resultant power spectrum should be unchanged when we conduct a random sampling equally 
to centrals and satellites, and the only effect is an amplified shot noise, because this operation is a Poisson sampling.
After some tests, we find that we cannot find a better fit when we randomly discard satellite subhalos while
assigning mock LRGs to all the centrals. We thus conduct random sampling only to the central population.
The probability of a central subhalo to host a mock LRG, $p_{\rm cen}$, is the third parameter of this model.

\item{\bf Model 5: centrals $+$ satellites, cosmology varied}\mbox{}\\
Finally, we fit the observed power spectrum varying the cosmological parameters.
We adopt a model similar to the baseline model, Model 2. 
In addition to the two parameters in Model 2, we simultaneously vary three parameters that will be 
introduced shortly in Section~\ref{subsec:RSD}. We investigate with this model the robustness
of the result obtained with Model 2 for the parameters that characterize the properties of LRGs. At the same time,
we discuss the prospects to simultaneously determine the cosmological parameters together with the model parameters
in Model 2, and the possible systematic bias in the derived cosmological parameters when we
misunderstand the nature of galaxies.
\end{description}

\subsection{Determining the model parameters with the Markov-chain Monte Carlo method}
\label{subsec:fit}
For a given mock LRG catalog specified with the model parameters described above,
we measure the multipole moments of the power spectrum, evaluate the goodness of fit to the observed data 
and finally put constraints on the parameters as described in this section. 

We first explain the measurement of the multiple moments of the power spectrum.
We adopt the distant observer approximation and displace the mock LRGs as
\be
\mathbf{s} = \mathbf{x} - f\,u_z\hat{z},
\label{eq:map_rsd}
\ee
where $\mathbf{x}$ and $\mathbf{s}$ respectively denote the position of a mock LRG in real and redshift space,
$u_z = -v_z/(aHf)$ is the $z$ component of the normalized velocity and $\hat{z}$ is the unit vector along the line of sight. 
We then construct a density field in redshift space on a regular lattice with $1,024^3$
grid points using the nearest-grid-point (NGP) interpolation scheme \citep{Hockney81}. 
We have 11 density fields, $\delta^{(n)}$, from 11 random realizations given the model parameters, 
where $n=1,\dots,11$. After applying the Fast Fourier Transformation (FFT) to $\delta^{(n)}$, we correct for the window 
effect arising from the NGP interpolation by dividing $\delta^{(n)}_\mathbf{k}$ by an appropriate window kernel 
(e.g., \citealt{Jing05}).
We record the square of the density contrast multiplied by an appropriate weight,
\be
\hat{P}_\ell^{(n)}(\mathbf{k}) \equiv \frac{2\ell+1}{2}\mathcal{P}_\ell(\mu_\mathbf{k})\left|\delta_\mathbf{k}^{(n)}\right|^2,
\label{eq:unbinned}
\ee
for $\ell = 0$, $2$ and $4$. Note that we subtract the shot noise contribution, the inverse of the number density of the
mock LRGs, from the monopole moment.
We store all the modes up to $k<0.31\,h\,\mathrm{Mpc}^{-1}$, and the total number of available modes
from the $11$ realizations is about $(3.77\times10^5)\times11\simeq4.15\times10^6$ in Models 1, 2, 3 and 4 
where we do not consider the Alcock-Paczynski distortion, and it changes with the model parameters in Model 5.

Once a set of $\hat{P}_\ell$ is measured, we compare it with the observed multipole moments.
We fit $\hat{P}_\ell$ with the cubic B-spline function (note that this procedure is not an interpolation but a fit) to
have smooth predictions of the multipole moments of the power spectrum and evaluate them at the exact
wavenumbers where observational data are available. The breakpoints to construct the B-spline function are chosen
so as not to smooth the multipoles too much to erase the feature of BAOs (see Appendix~\ref{sec:quality}).
This procedure greatly reduces the artificial effect arising from the discrete sampling of modes in Fourier space 
along the $\mu_\mathbf{k}$ direction, which adds a significant noisy pattern depending on the box size of the simulation
especially to the higher multipoles (i.e., $\ell = 2$ and $4$; see \citealt{Nishimichi11,Taruya12}).
See Appendix~\ref{sec:quality} for more detail.

After the multipole moments of the power spectrum are obtained, we evaluate the goodness of fit defined as
\be
\chi^2 \equiv \sum_{\ell=0,2,4}\sum_{k<k_\mathrm{max}}\left[\frac{P_{\ell,\mathrm{mock}}(k)-P_{\ell,\mathrm{obs}}(k)}
{\Delta P_{\ell,\mathrm{obs}}(k)}\right]^2,
\ee
where $P_{\ell,\mathrm{mock}}$ and $P_{\ell,\mathrm{obs}}$ respectively denote the simulated and observed power 
spectra, and $\Delta P_{\ell,\mathrm{obs}}$ is the statistical error (i.e., the standard deviation) on the observed spectrum.
We simply neglect the statistical error on the model power spectrum because it is small (typically $10$ per cent of 
$\Delta P_{\ell,\mathrm{obs}}$; see Appendix~\ref{sec:quality}).

We search for the parameter set that gives the smallest $\chi^2$ 
employing a Markov-chain Monte Carlo (MCMC) method.
Since we have to perform the FFT to the 11 density fields and subsequently the B-spline fitting to a lot of data points
($\sim\mathcal{O}(10^6)$) at each chain of the MCMC, the computational cost is quite expensive.
It takes about one minute with a processor (16 cores) in the Cray XC30 cluster
to evaluate the power spectrum for a given model parameter, and we need
up to $100,000$ chains to have a converged result depending on the model. 
We can do this analysis in about one week using $10$ processors parallelly.

\subsection{Simulating the redshift-space distortions}
\label{subsec:RSD}
If one wishes to constrain cosmological parameters, one may need to prepare
theoretical predictions for the whole parameter space of interest. 
This is computationally unfeasible with $N$-body simulations and, as a matter of fact, we 
have a converged prediction of the power spectrum only for one cosmological model.
However, we can still simulate the anisotropy of the clustering expected for different cosmologies 
using that simulation dataset for one particular cosmological model. 

We start with the positions and velocities of mock LRGs given a set of model parameters
described in Section~\ref{subsec:model}.
We then introduce one parameter that controls the amplitude of RSDs.
We denote it by $\alpha_\mathrm{v}$, and we map the positions of mock LRGs from real to redshift space as
\be
\mathbf{s} = \mathbf{x} - \alpha_\mathrm{v}\,f\,u_z\hat{z},
\ee
instead of equation~(\ref{eq:map_rsd}). The distribution of mock LRGs is still given in a cube up to here.
We then deform this periodic cube according to two parameters, $\alpha_{\parallel}$ and $\alpha_\perp$.
Namely, we stretch the simulation box by a factor of $\alpha_{\parallel}$ ($\alpha_{\perp}$) along the line-of-sight 
(plane-of-sky) direction under the distant observer approximation.
We then measure the multipoles of the power spectrum with the same procedure as before, but from 
$11$ rectangular cuboids instead of $11$ cubes.

The parameter $\alpha_\mathrm{v}$ carries information about the growth-rate as explained in what follows.
Adopting a theoretical template based on General Relativity (GR), 
one usually treats the parameter combination 
$f(z)\sigma_8(z)$ as an independent free parameter that may differ from the value for GR. 
This combination controls the amplitude of the linear velocity field, and
can be wavenumber-dependent in some modified gravity scenarios such as the $f(R)$ gravity 
\citep{Hu:2007lr,Starobinsky:2007fk}.
In practice, a constant $f\sigma_8$ is more tractable and thus usually adopted in the literature. 
What we do with the parameter $\alpha_\mathrm{v}$ is the {\it nonlinear} analogy of this procedure.
By selecting a value of $\alpha_\mathrm{v}$ different from unity, we change the nonlinear velocity field from what 
is expected for GR. On sufficiently large scales where nonlinear contamination is negligible, 
this parameter can be identified with the following combination:
\be
\alpha_\mathrm{v} = \frac{(f\sigma_8)(z=0.3)}{(f\sigma_8)_\mathrm{fid}(z=0.35)}.
\label{eq:cosmo1}
\ee
Note that we take into account in the above a small difference of the redshift between the survey ($z=0.3$) and 
the simulation ($z=0.35$). 
Here and hereafter, variables with a subscript ``fid" refer to those
assumed in the simulations or used in the redshift-distance relation when the power spectrum is measured.

The AP distortion induced to the simulations with our procedure is based on the mapping of wavevectors
under the distant observer approximation \citep{Matsubara:1996qy,Ballinger96}:
\be
P_\mathrm{obs}(\bfk) = [\alpha_\parallel \alpha_\perp^2] P_\mathrm{true}(\bfq),\nonumber\\
q_\parallel = \alpha_\parallel k_\parallel,\qquad q_\perp = \alpha_\perp k_\perp,
\label{eq:AP}
\ee
where $\bfq$ is the wavevector in the unknown true coordinate system with $q_\parallel$ and $q_\perp$ respectively
being its components parallel and perpendicular to the line of sight, while the observed coordinate is $\bfk$ 
with $k_\parallel, k_\perp$ defined analogously.
The prefactor in the square bracket represents the change in the volume element by the coordinate transformation. 
This mapping formula derived for analytical models of 
the power spectrum is automatically realized by measuring the power spectrum from the deformed simulation boxes.

With a help of the BAO feature clearly visible in the observed spectrum, we can conduct a geometrical test.
The parameter $\alpha_\parallel$ is related to the comoving distance along the line of sight, and thus reflects
the ratio of the Hubble parameter between the true unknown cosmology and the one assumed when we convert the redshifts
into comoving distances. In addition, the parameter $\alpha_\parallel$ depends on the 
acoustic horizon scale at the baryon drag epoch \citep{Eisenstein98},
since the true acoustic scale might be different from the one realized in the simulations.
We denote the acoustic scale at the drag epoch by $r_\mathrm{s}$, and compute it with the \textsc{CAMB} code.
Indeed, this quantity computed for our fiducial cosmological model is larger than that for the PLANCK cosmology 
by about five per cent when measured in units of $h^{-1}\mathrm{Mpc}$ (this difference mainly comes from the difference in 
$h$; see Table~\ref{tab:cosmo}).
Taking this difference into account, we have
\be
\alpha_\parallel = \frac{H(z)r_\mathrm{s}}{H_\mathrm{fid}(z)r_\mathrm{s, fid}},
\label{eq:cosmo2}
\ee
at the effective redshift of the measured power spectrum, $z=0.3$.
Similarly, the angular diameter distance, $D_\mathrm{A}$,
can be constrained through the parameter $\alpha_\perp$:
\be
\alpha_\perp = \frac{D_\mathrm{A, fid}(z)r_\mathrm{s}}{D_\mathrm{A}(z)r_\mathrm{s, fid}},
\label{eq:cosmo3}
\ee
at $z=0.3$.

We simultaneously vary $\alpha_\mathrm{v}$, $\alpha_\parallel$ and $\alpha_\perp$ as well as 
$M_\mathrm{host, min}$ and $M_\mathrm{sub, min}$ to find the best-fit parameter set for Model 5.
We discuss the robustness of the constraints on the parameters for the mock LRGs 
when the cosmological assumptions are relaxed by comparing the results of Model 2 and 5. 
Also, we show the derived cosmological constraints and
compare them with those in the literature to demonstrate the prospect of analyzing observational
data with theoretical predictions from simulations instead of analytical models.

\section{Results}
\label{sec:results}
Now we are in position to show the results of the MCMC analysis explained so far.
We first discuss the importance of the satellite population to model the anisotropic clustering of LRGs
by showing the results of the fit with Models 1 and 2 in Section~\ref{subsec:subhalo}. 
We then compare the satellite fraction derived with Model 2 and that
with Models 3 and 4 where centrals and satellites host mock LRGs with different criteria 
in Section~\ref{subsec:extension}. 
We further discuss the robustness of the results against cosmological uncertainties in Section~\ref{subsec:robust}. 
Some cosmological implications are given in Section~\ref{subsec:cosmo}.
We finally compare the multiplicity function of our best-fit models with observation in Section~\ref{subsec:multi}.
The best-fit model parameters as well as the goodness of fit are summarized in Table~\ref{tab:best}.

\begin{table*}
\begin{minipage}{168mm}
\begin{center}
\caption{Ten best-fit models and their goodness of fit. Masses are given in unit of $10^{13}\,h^{-1}M_\odot$, while
the number density $n_\mathrm{g}$ is in $10^{-4}\,h^3\,\mathrm{Mpc}^{-3}$. The first column shows the name of the 
model. The best-fit model parameters are shown in the second to the sixth column. The mean number density
$n_\mathrm{g}$, the fraction of satellite LRGs $f_\mathrm{sat}$ and the fraction $f_\mathrm{sat}^*$ 
when a different definition of satellites are adopted (see text) are listed in the seventh to the eleventh column.
We finally show the chi-squared statistics and the figure number in which the multipoles are shown.\label{tab:best}
}
\scalebox{0.8}{
\begin{tabular}{l ccccc ccccc cccc c}
\hline
Model & $M_\mathrm{host,min}$ & $M_\mathrm{sub,min}$ & $M_\mathrm{sat,min}$ & $M_\mathrm{cen,min}$ & 
$p_\mathrm{cen}$ & $n_g$ & $f_\mathrm{sat}$ & $\pm(1\sigma)$ & $\pm(2\sigma)$ & $f_\mathrm{sat}^*$ &
$\chi^2_\mathrm{0,min}$ & $\chi^2_\mathrm{2,min}$ & $\chi^2_\mathrm{4,min}$ 
& $\chi^2_\mathrm{red}$ & $P_\ell(k)$ \\ \hline
1a & $2.50$ & - & - & - & - & $1.28$ & $0$ & & & $0$ & $109.2$ & $916.8$ & $27.1$ & $11.8$ & Fig.~\ref{fig:P1} \\ 
1b & $2.70$ & - & - & - & - & $1.16$ & $0$ & & & $0$ & $119.0$ & $656.5$ & $34.6$ & $9.10$ & \\ \hline
2a & $1.29$ & $0.211$ & - & - & - & $4.07$ & $0.300$ & $^{+0.017}_{-0.012}$ & $^{+0.032}_{-0.024}$ & $0.098$ &
$48.0$ & $28.4$ & $27.8$ & $1.18$ & Fig.~\ref{fig:P2} \\ 
2b & $1.45$ & $0.299$ & - & - & - & $3.30$ & $0.248$ & $^{+0.015}_{-0.010}$ & $^{+0.029}_{-0.024}$ & $0.093$ &
$31.2$ & $28.7$ & $30.7$ & $1.03$ & \\ \hline
3a & $1.29$ & - & $0.216$ & $1.27$ & - & $3.76$ & $0.316$ & $^{+0.009}_{-0.019}$ & $^{+0.023}_{-0.033}$ & $0.099$ &
$39.9$ & $28.6$ & $26.3$ & $1.09$ & Fig.~\ref{fig:P3} \\
3b & $1.47$ & - & $0.316$ & $1.46$ & - & $2.97$ & $0.259$ & $^{+0.009}_{-0.019}$ & $^{+0.024}_{-0.033}$ & $0.094$ &
$26.7$ & $29.2$ & $29.1$ & $0.976$ & \\ \hline
4a & $0.981$ & $0.886$ & - & - & $0.149$ & $0.740$ & $0.260$ & $^{+0.020}_{-0.006}$ & $^{+0.036}_{-0.019}$ & $0.073$ & 
$26.8$ & $26.8$ & $21.6$ & $0.865$ & Fig.~\ref{fig:P4} \\
4b & $1.16$ & $1.13$ & - & - & $0.167$ & $0.622$ & $0.213$ & $^{+0.019}_{-0.008}$ & $^{+0.038}_{-0.020}$ & $0.060$ &
$21.7$ & $27.7$ & $22.9$ & $0.830$ & \\ \hline
5a & $1.81$ & $0.427$ & - & - & - & $2.85$ & $0.214$ & $^{+0.017}_{-0.030}$ & $^{+0.046}_{-0.050}$ & $0.089$ &
$21.6$ & $31.1$ & $26.4$ & $0.930$ & Fig.~\ref{fig:P5} \\
5b & $1.87$ & $0.465$ & - & - & - & $2.67$ & $0.201$ & $^{+0.018}_{-0.027}$ & $^{+0.046}_{-0.049}$ & $0.088$ &
$20.5$ & $29.5$ & $30.8$ & $0.950$ & \\
\hline
\end{tabular}}
\end{center}
\end{minipage}
\end{table*}%

\subsection{Importance of satellites}
\label{subsec:subhalo}

\begin{figure}
\begin{center}
\includegraphics[height=7.4cm]{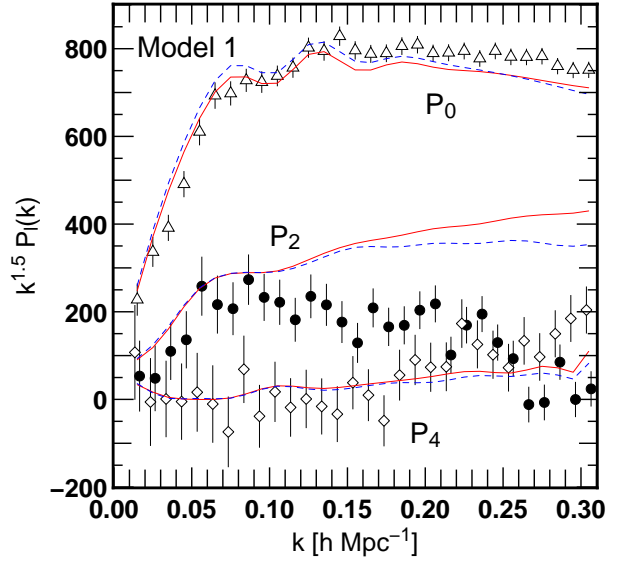}
\caption{Best-fit power spectra against observation assuming one-to-one correspondence between
LRGs and halos (Model 1).
Symbols show the observed multipoles of the power spectrum 
(open triangles: monopole, filled circles: quadrupole and open diamonds: hexadecapole).
The solid (dashed) curves depict the result of Model 1a (b).
Note that symbols for the quadrupole (hexadecapole) moment are horizontally offset by $+0.0015$ ($-0.0015$)
to avoid overlap.}
\label{fig:P1}
\end{center}
\end{figure}

We start this section by showing the result of Model 1.
The best-fit multipoles of the power spectrum are shown in Fig.~\ref{fig:P1} together with the observation.
The results of Model 1a and 1b are respectively plotted in solid and dashed lines.
It is clear from the figure that we cannot simultaneously fit the three multipoles with the central population alone.
There exists a mismatch between the observed and the model quadrupole at $k\simgt0.1\,h\,{\rm Mpc}^{-1}$.
The difference between the two models (Model 1a and 1b) is visible, but is much smaller than
the discrepancy between the observation and the two models.
Also, we can observe a difference in the broadband shape of the monopole moment. Since we have only one
parameter in Model 1 (i.e., the minimum mass of the host halos), it might be difficult to simultaneously adjust the 
amplitude and the shape of the spectrum. 

To be more quantitative, we show the minimum value of $\chi^2$ in Table~\ref{tab:best}.
The values for the three multipoles, $\chi^2_{\ell, \mathrm{min}}\,(\ell = 0,\,2, \,\mathrm{and}\,4)$ 
and their sum divided by the degree of freedom (the reduced chi-squared, $\chi^2_\mathrm{red}$) are listed in the table. 
Note that each multipole moment has $30$ data points up to $k_\mathrm{max} = 0.305\,h\,\mathrm{Mpc}^{-1}$.
Clearly, both Model 1a and 1b give a poor fit to the observed data with Model 1b being slightly better 
($\chi^2_\mathrm{red}=11.8$ and $9.10$, respectively), 
and most of the discrepancy comes from the quadrupole moment.

\begin{figure}
\begin{center}
\includegraphics[height=7.4cm]{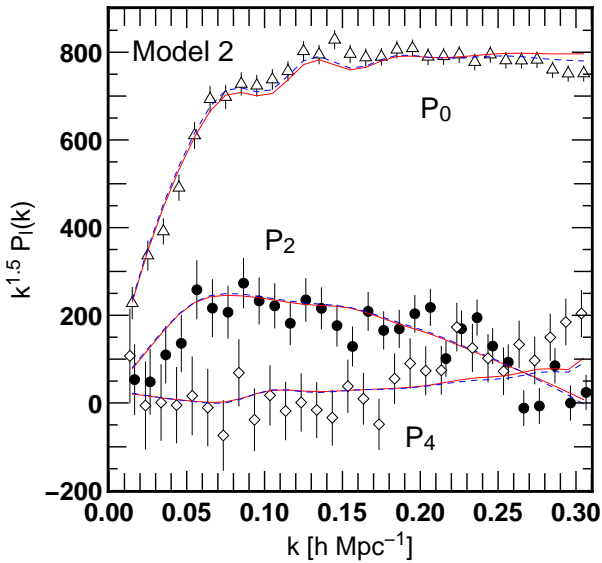}
\caption{Same as in Fig.~\ref{fig:P1}, but for Model 2.}
\label{fig:P2}
\end{center}
\end{figure}

\begin{figure}
\begin{center}
\includegraphics[height=3.8cm]{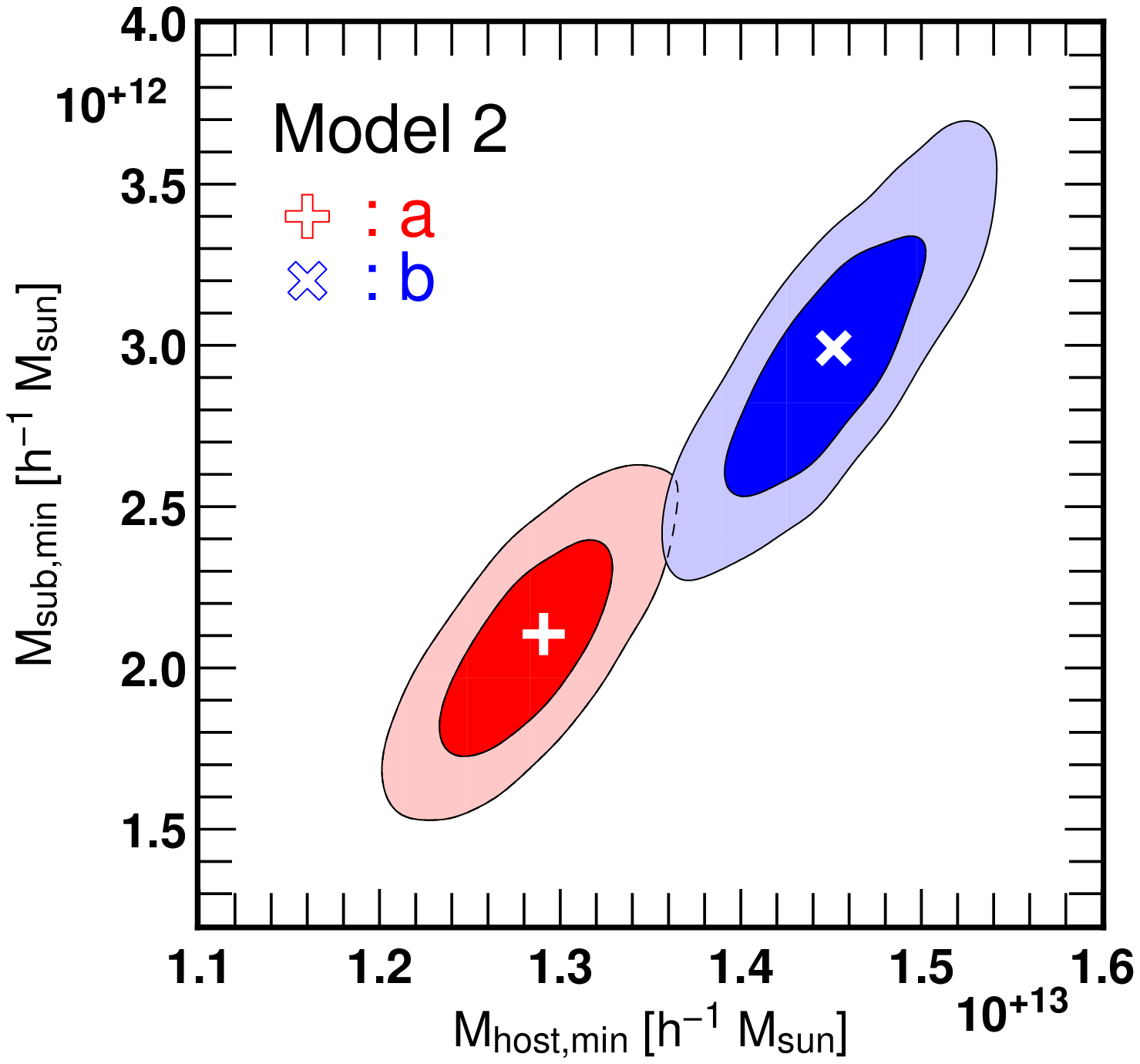}
\includegraphics[height=3.8cm]{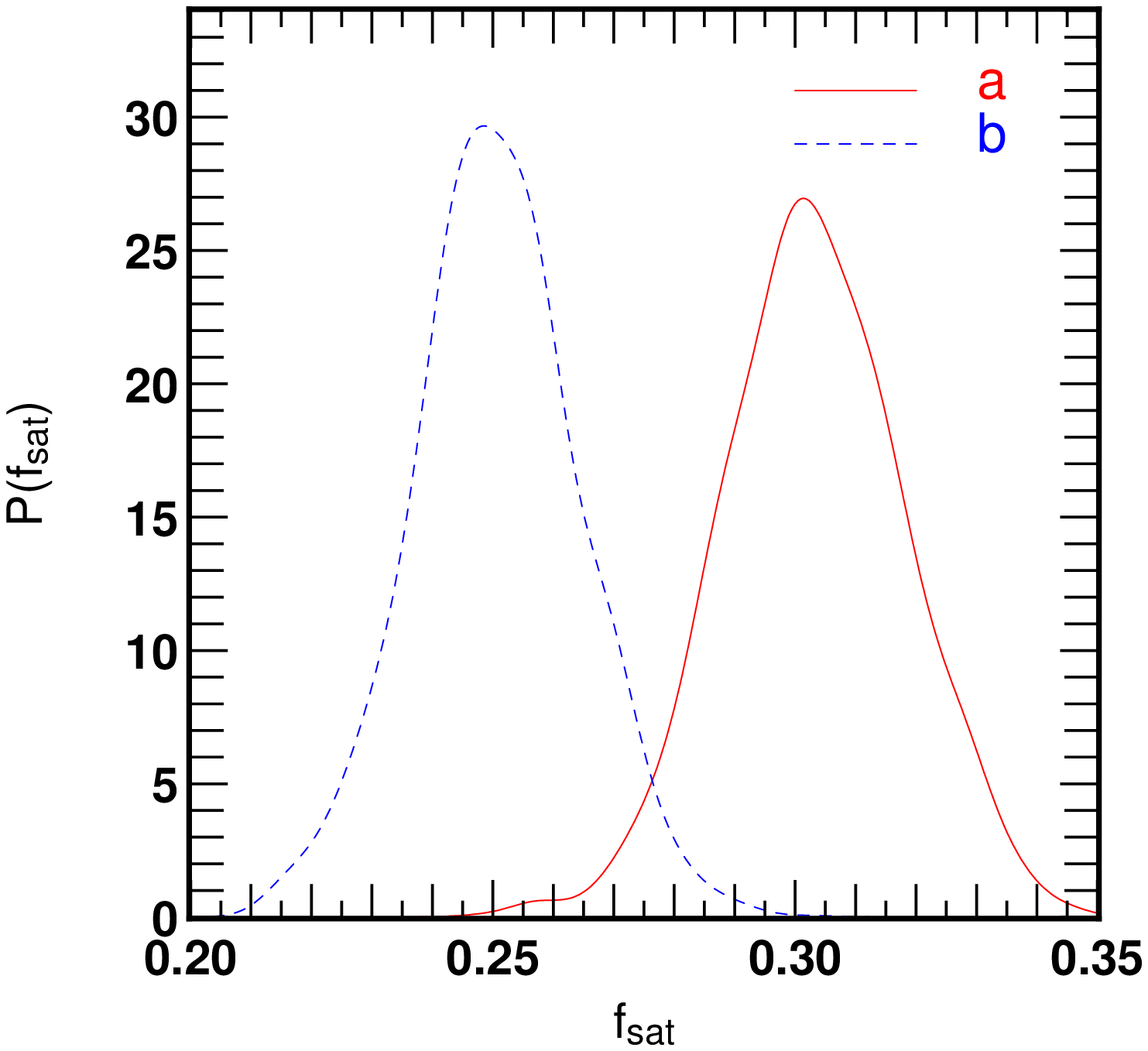}
\caption{Constraints on the parameters of Model 2.
We plot $1$ and $2\sigma$ allowed regions for parameters characterizing the mock LRGs (left), and
the posterior distribution of the fraction of satellites (right). 
In the left panel, Model 2a is shown by red contours while Model 2b is plotted with blue.
We depict by solid (dashed) line the result of Model 2a (2b) in the right panel.}
\label{fig:M2}
\end{center}
\end{figure}

We next discuss the result of Model 2 in which we assign mock LRGs to both centrals and satellites. 
We can see in Fig.~\ref{fig:P2} a substantial improvement of fit over Model 1 plotted in Fig.~\ref{fig:P1}.
The values of $\chi^2_\mathrm{red}$ are $1.18$ and $1.03$ for Model 2a and 2b, respectively, 
implying that the modeling is sufficient to explain the observed multipole moments up to $k\simeq0.3\,h\,\mathrm{Mpc}^{-1}$.
The constraints on the model parameters are shown in the left panel of Fig.~\ref{fig:M2} with the best-fit values
indicated by the plus (cross) symbols for Model 2a (2b).
The two parameters, $M_\mathrm{host,min}$ and $M_\mathrm{sub,min}$,
show a positive correlation in both models, and the allowed regions for the two models are significantly different.

We also plot in the right panel the constraint on the fraction of LRGs hosted by satellite subhalos, $f_\mathrm{sat}$.
A satellite fraction of $f_\mathrm{sat}\sim0.3$ is favored by the observed power spectrum 
for Model 2a, while a smaller value is derived for Model 2b ($f_\mathrm{sat}\sim0.25$; dashed). 
The difference between Models 2a and 2b mainly comes from the different velocity dispersion of mock LRGs.
We have seen that Model 1 can not explain the suppress of the quadrupole at $k\simgt0.1\,h\,\mathrm{Mpc}^{-1}$.
By adding a significant fraction of satellite LRGs, Model 2 can explain the quadrupole thanks to a larger velocity 
dispersion of the mock LRGs in satellite subhalos than those in centrals.
Furthermore, since the mock LRGs in Model 2b have a larger velocity dispersion than in Model 2a, 
the former needs a smaller fraction of satellite LRGs to match the observed power spectrum. 
The result of this section highlights the importance of satellite LRGs to
understand the observed anisotropic clustering.
Since the chi-squared values are very similar in these two models, we conclude that it is difficult to constrain
the additional velocity contribution in Model 2b from the large scale analysis alone.
This component is degenerate with the fraction of satellites, and thus we can put a constraint only on the 
total velocity dispersion of centrals and satellites, which determines the magnitude of the Fingers-of-God effect.

Our constraint on the fraction of satellite LRGs might be counterintuitive given the small fraction of 
multiple-LRG systems suggested from observations ($\sim 5$ per cent).
Indeed, the lower bound on this parameter is given as $f_\mathrm{sat} > 0.27$ ($0.22$) at $95$ per cent C.L. in Model 2a (2b)
from our analysis, while the value is constrained to $6.36^{+0.38}_{-0.39}$ per cent ($95$ per cent C.L.) by \citet{Reid09} 
based on the observed multiplicity function of LRGs.
As discussed in more detail below, this apparent discrepancy in the satellite fraction
partly comes from different definitions of satellites. Also, our result suggests that the central subhalos in our terminology
do not always host a LRG. We reserve a thorough discussion on the compatibility of our mock LRGs until
Section~\ref{subsec:multi}, and keep testing the robustness of our constraints against different assumptions
in the model of LRGs in what follows.
 
\subsection{Extension of the simplest model}
\label{subsec:extension}
Now the important question to ask here is whether the seemingly large satellite fraction obtained in the previous
section using Model 2 is real or just a consequence of incorrect modeling of LRGs.
We here adopt Model 3 and 4 and allow additional parameters to float.
Since these parameters control the selection criteria of satellites and centrals individually, we can
check the robustness of the satellite fraction derived in the previous section.

\begin{figure}
\begin{center}
\includegraphics[height=7.4cm]{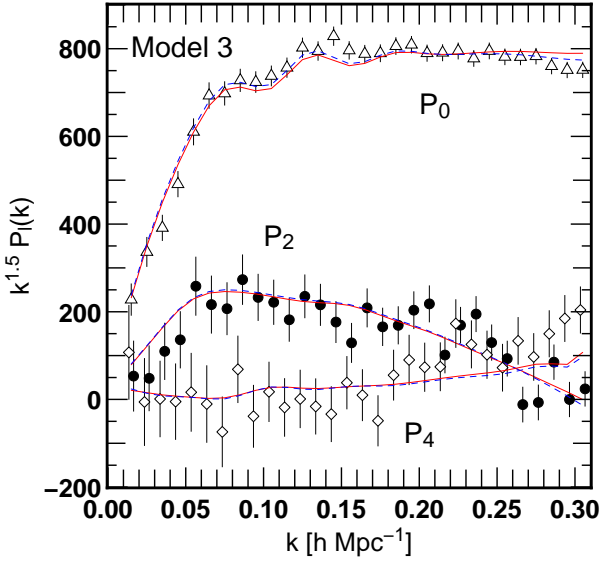}
\caption{Same as in Fig.~\ref{fig:P1}, but for Model 3.}
\label{fig:P3}
\end{center}
\end{figure}

\begin{figure}
\begin{center}
\begin{minipage}{0.47\columnwidth}
\includegraphics[height=3.8cm]{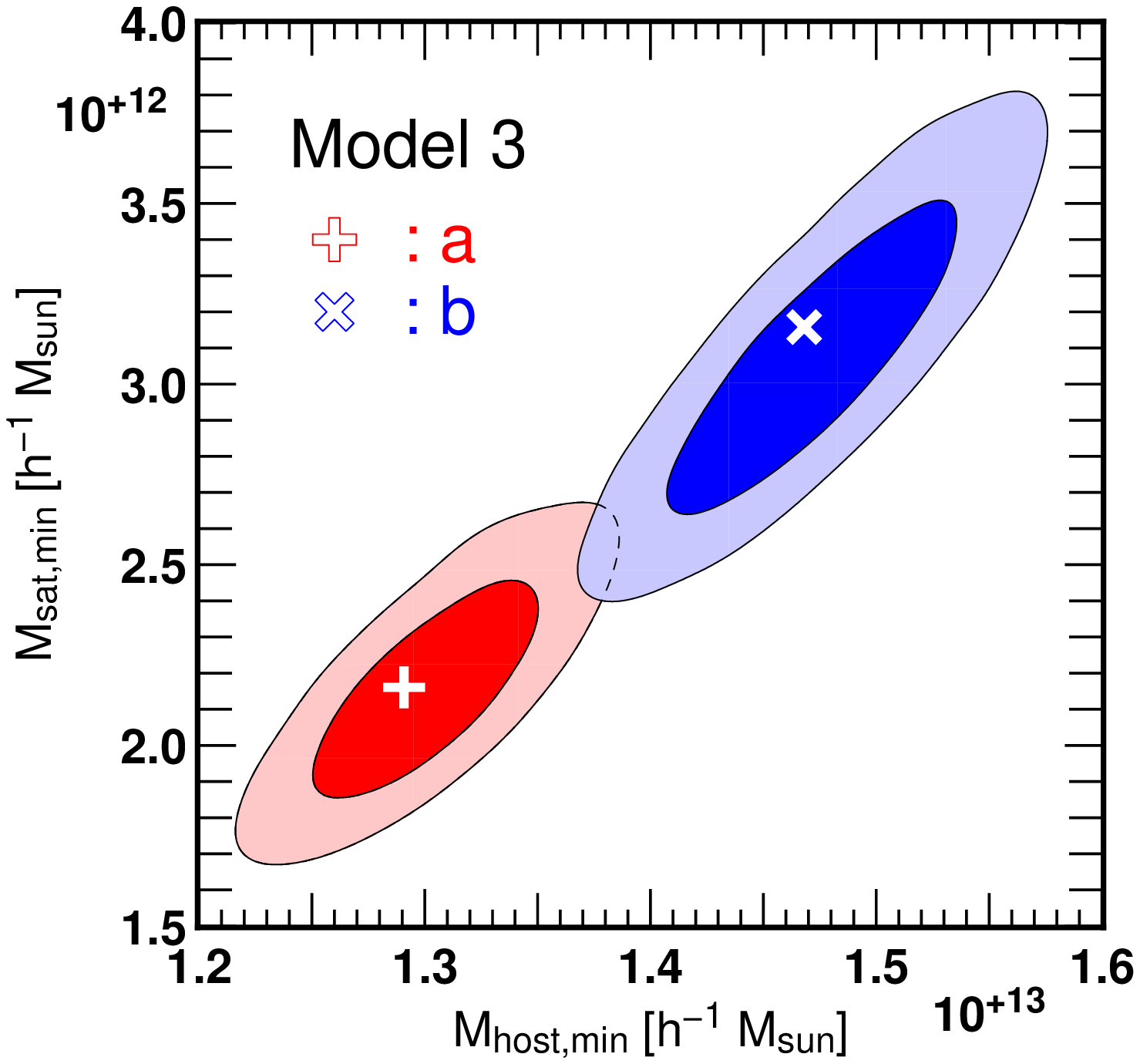}
\end{minipage}
\begin{minipage}{0.47\columnwidth}
\includegraphics[height=3.8cm]{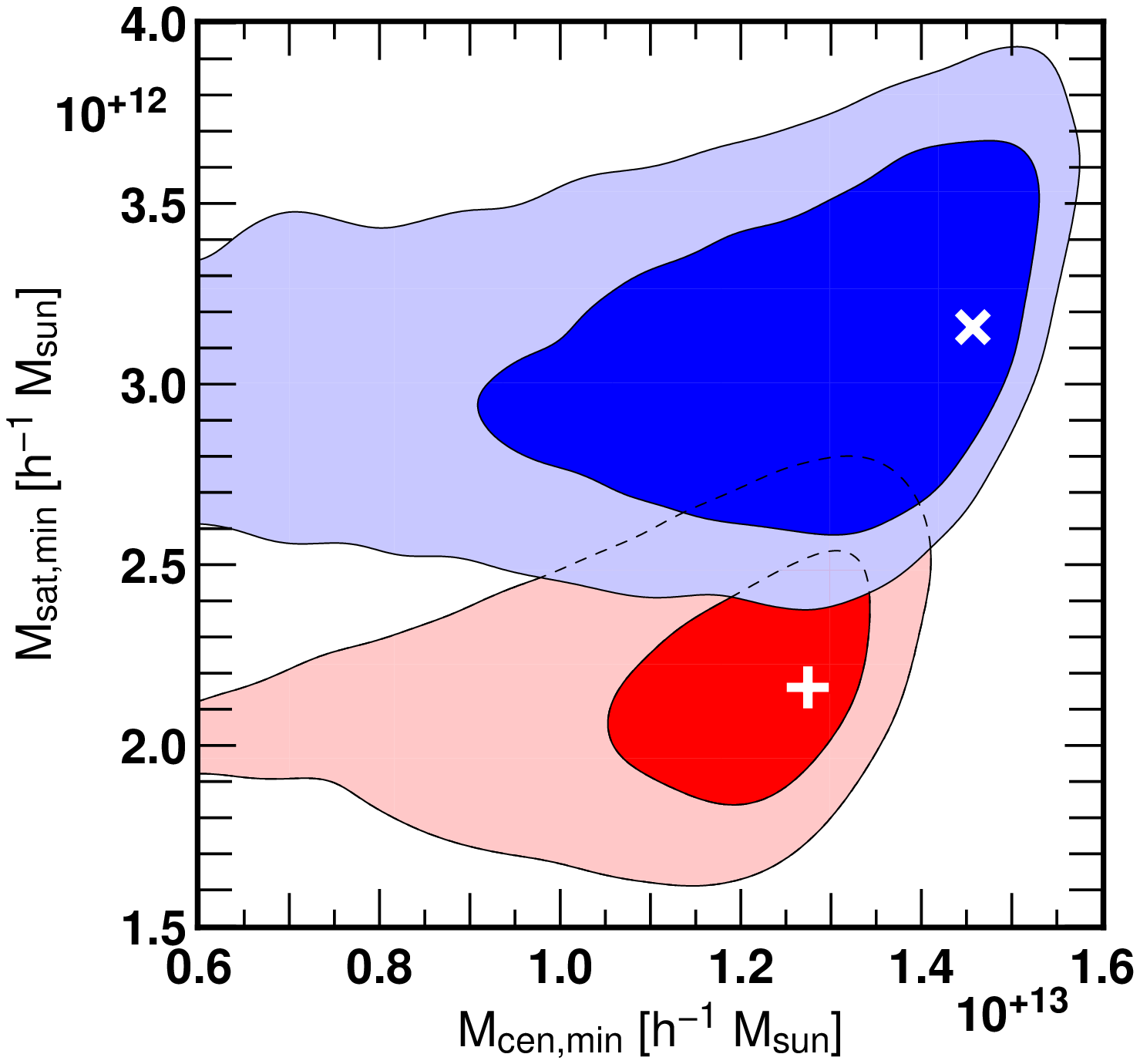}
\end{minipage}
\end{center}
\begin{center}
\begin{minipage}{0.47\columnwidth}
\includegraphics[height=3.8cm]{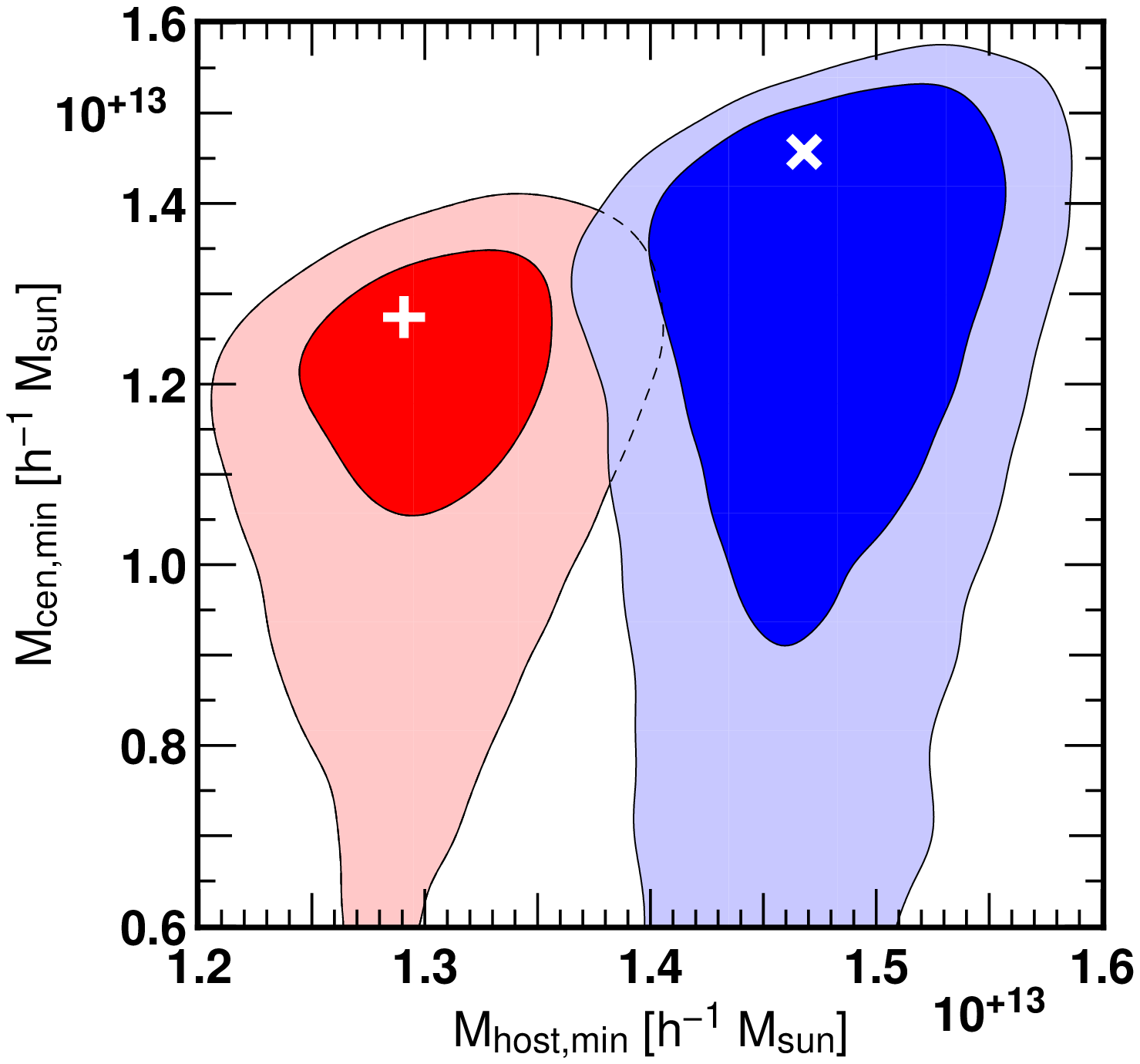}
\end{minipage}
\begin{minipage}{0.47\columnwidth}
\includegraphics[height=3.8cm]{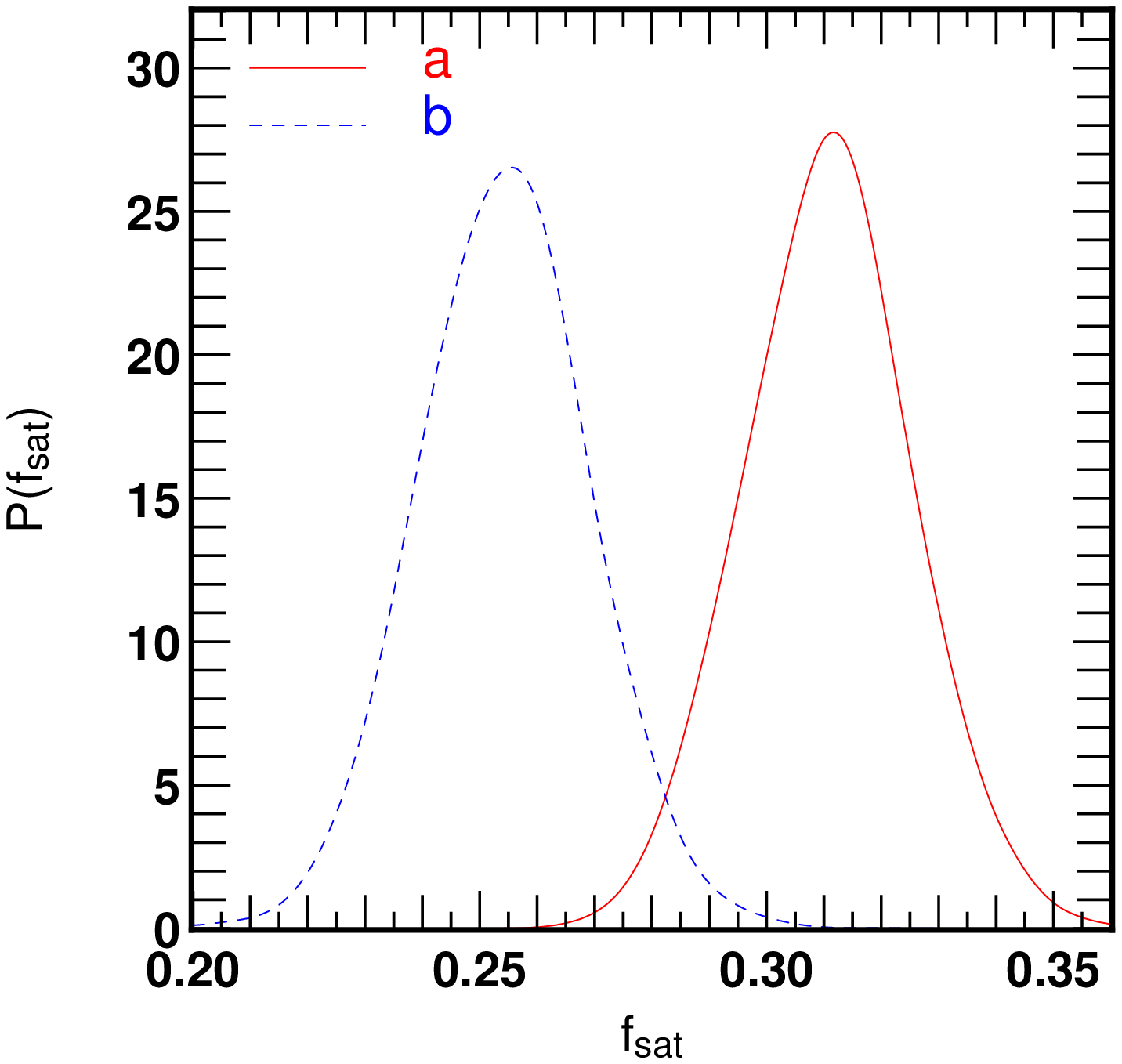}
\end{minipage}
\caption{Same as in Fig.~\ref{fig:M2}, but for Model 3.}
\label{fig:M3}
\end{center}
\end{figure}

The result of Model 3 is plotted in Figs~\ref{fig:P3} and \ref{fig:M3} for the best-fit power spectrum and
the constraints on the model parameters, respectively. It is difficult to judge by eye, but the fit is slightly improved 
over Model 2 as the values of $\chi^2_\mathrm{red}$ suggest (see Table~\ref{tab:best}). 
The constraints in $M_\mathrm{host,min}$-$M_\mathrm{sat,min}$ plane 
shown in the top-left panel of Fig.~\ref{fig:M3} is very similar to those in the left panel of Fig.~\ref{fig:M2} 
for $M_\mathrm{host,min}$ and $M_\mathrm{sub,min}$.
This suggests that the satellite LRGs in Model 3 are almost the same as in Model 2. By contrast, we cannot put
a stringent constraint on the other parameter, $M_\mathrm{cen,min}$ (see top-right and bottom-left panel 
of Figure~\ref{fig:M3}). This is because given $M_\mathrm{host,min}$, there are only small number of
centrals having mass much smaller than $M_\mathrm{host,min}$. 
The best-fit parameter set, indicated by the plus and cross symbols, suggests a
large $M_\mathrm{cen,min}$ compared with $M_\mathrm{sat,min}$. Thus the slightly improved fit compared to
Model 2 comes from a reduced number of the central LRGs. As a result, the satellite fraction plotted in the bottom-right panel
of Fig.~\ref{fig:M3} shows a slightly larger value than in Model 2 (right panel of Figure~\ref{fig:M2}), 
but the difference is much smaller than the statistical uncertainty on that parameter.

\begin{figure}
\begin{center}
\includegraphics[height=7.4cm]{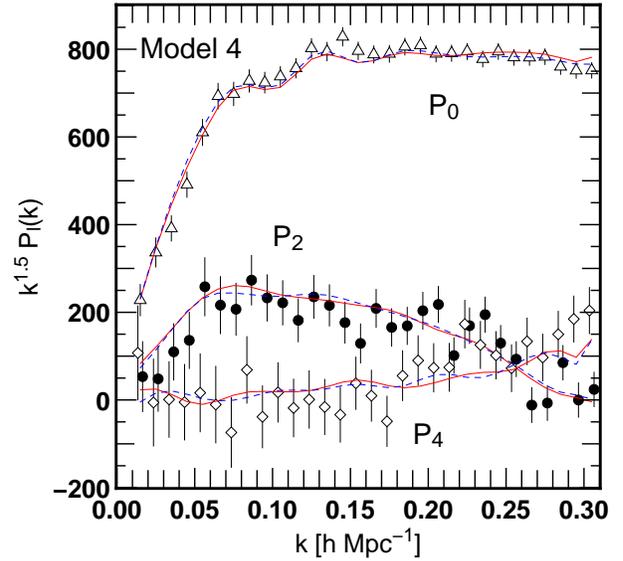}
\caption{Same as in Fig.~\ref{fig:P1}, but for Model 4.}
\label{fig:P4}
\end{center}
\end{figure}

Next, the best-fit power spectrum of Model 4 is shown in Fig.~\ref{fig:P4}.
Judging from the $\chi^2_\mathrm{red}$ listed in Table~\ref{tab:best}, this model gives a better fit
than Model 2 and 3 ($\chi^2_\mathrm{red} = 0.865$ and $0.830$ for Model 4a and 4b, respectively).
Although the values themselves should be used with caution because we ignored the off-diagonal components
of the covariance matrix, these values suggest a slight overfit to the observational data.
This might be explained as follows.
In Model 4, central subhalos are randomly selected with a probability $p_\mathrm{cen}$ to host LRGs.
Because of this random process, the resultant multipole moments can be different from one time to another
even one employs exactly the same model parameters. In other words, the increased shot noise in the model
spectrum, which is not taken into account in the MCMC analysis, can sometimes mimic the observed noise pattern
by chance and reduce the chi-squared statistics. Indeed, the best-fit multipoles are less smooth than in previous figures
because of this effect.

\begin{figure}
\begin{center}
\begin{minipage}{0.47\columnwidth}
\includegraphics[height=3.8cm]{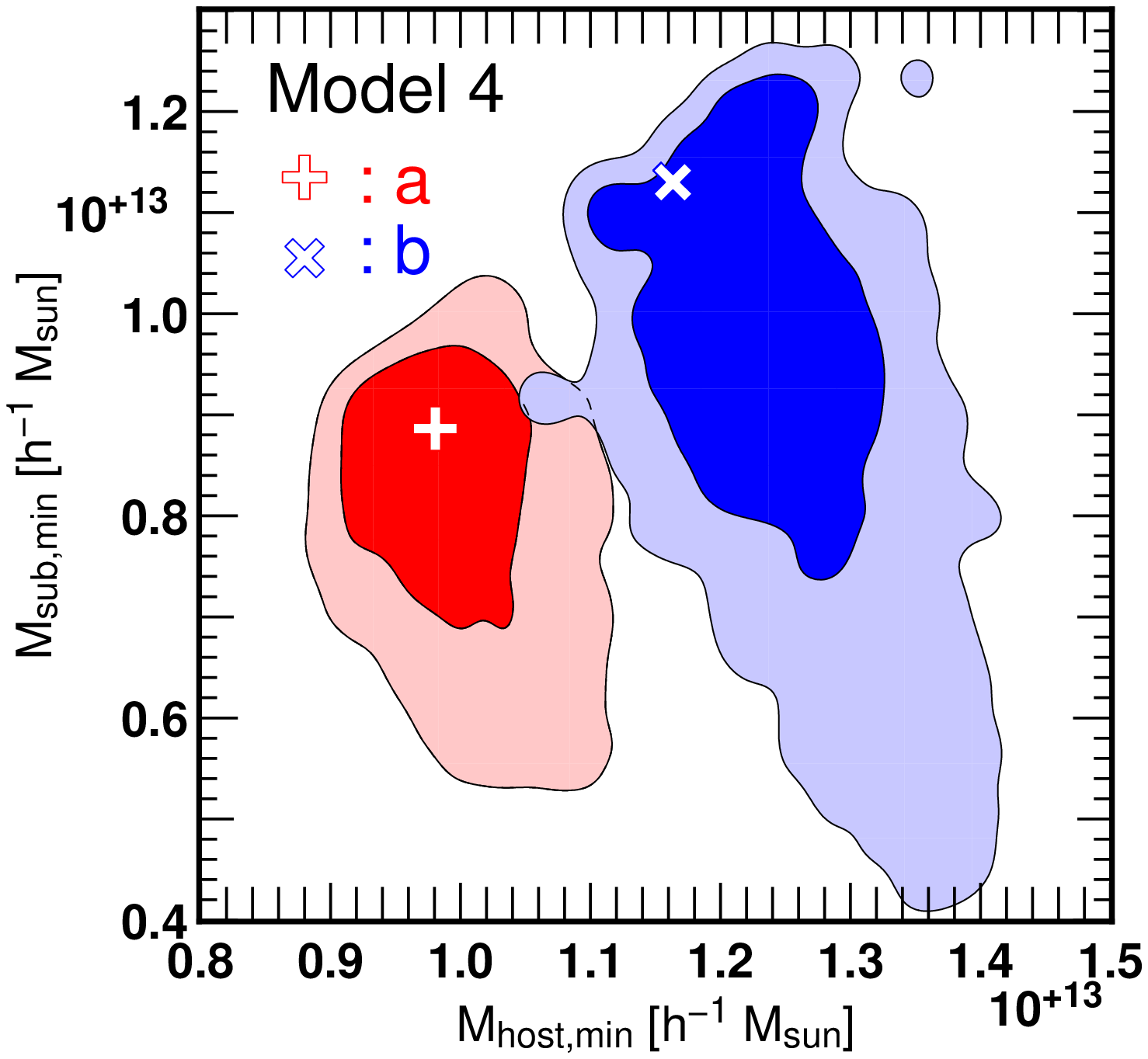}
\end{minipage}
\begin{minipage}{0.47\columnwidth}
\includegraphics[height=3.8cm]{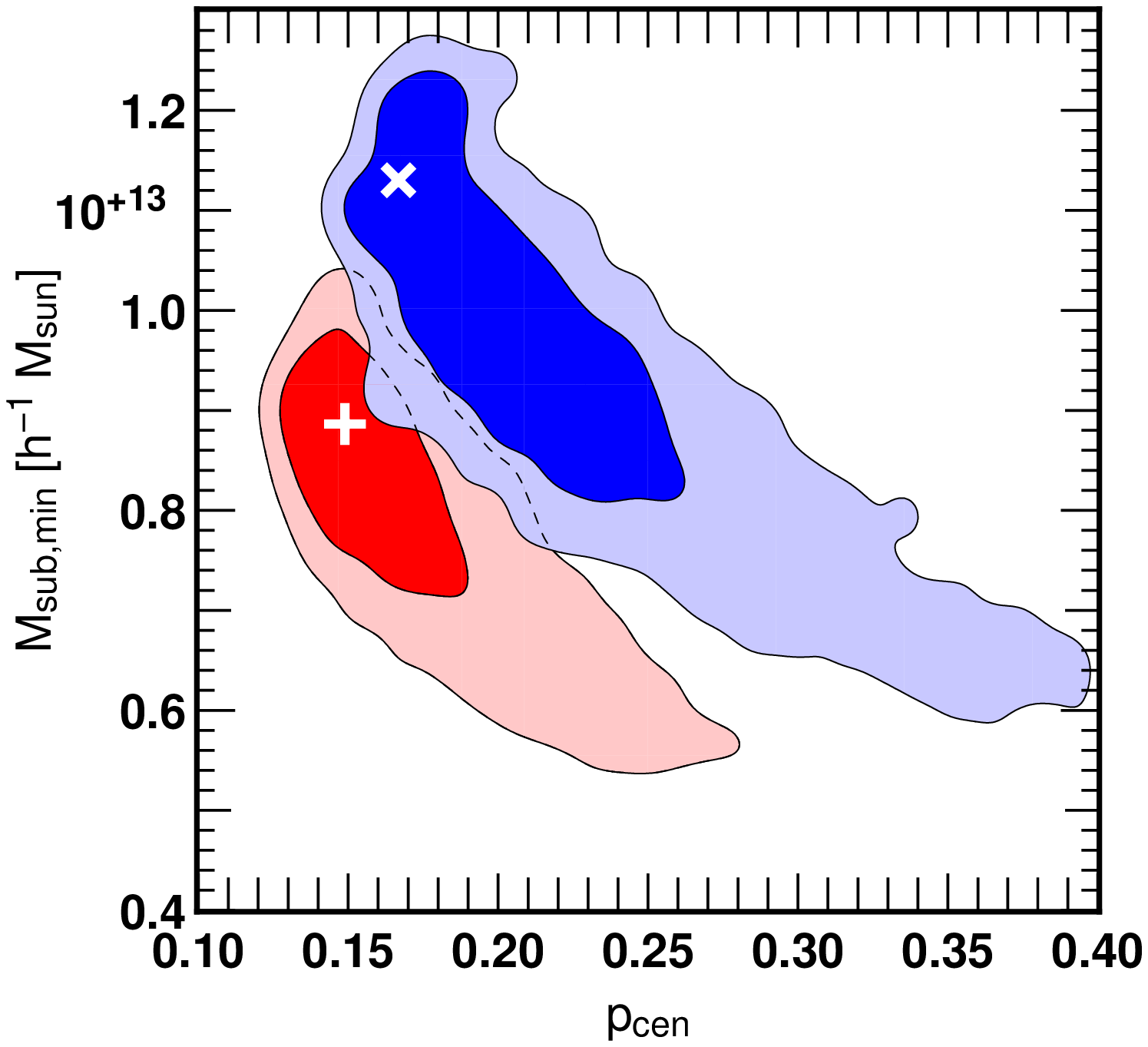}
\end{minipage}
\end{center}
\begin{center}
\begin{minipage}{0.47\columnwidth}
\includegraphics[height=3.8cm]{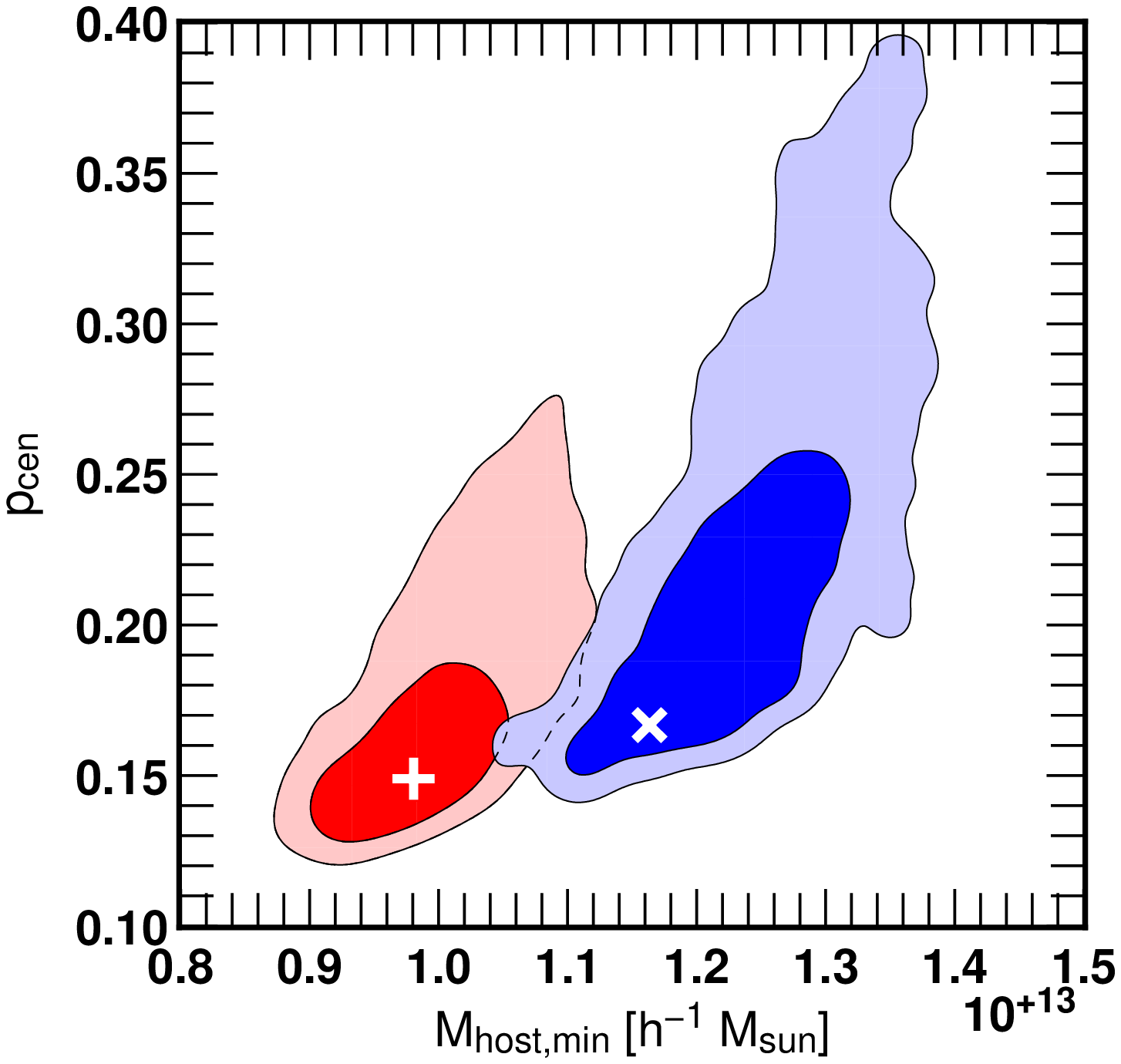}
\end{minipage}
\begin{minipage}{0.47\columnwidth}
\includegraphics[height=3.8cm]{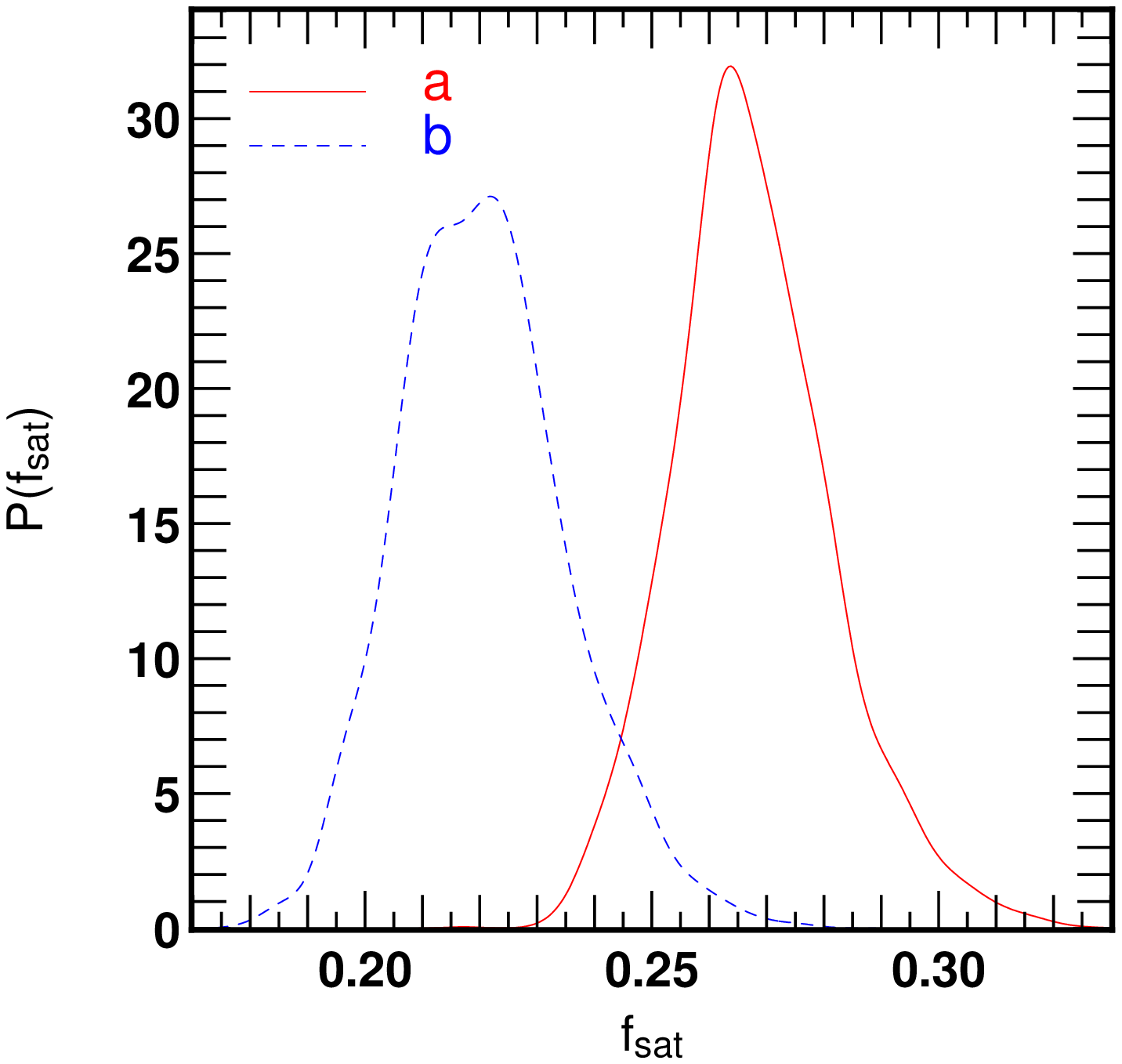}
\end{minipage}
\caption{Same as in Fig.~\ref{fig:M2}, but for Model 4.}
\label{fig:M4}
\end{center}
\end{figure}

The constraints on the model parameters are shown in Fig.~\ref{fig:M4}. Interestingly, the constraints
in the $M_\mathrm{host,min}$-$M_\mathrm{sub,min}$ plane shown in the top-left panel is significantly different from
that in Model 2 (Figure~\ref{fig:M2}): the minimum host halo mass is smaller, and the minimum subhalo
mass is larger in Model 4. Because of the smaller host halo mass, the number of centrals that pass the 
mass criterion increases, but it is then reduced by random sampling. 
The number of satellites is smaller because of a larger $M_\mathrm{sub,min}$,
and as a result, the mean number density of the final mock LRGs are much smaller than in Model 2 
(see Table~\ref{tab:best}). These differences might suggest two possible different nature of LRGs
inferred with Model 2 and with Model 4.
Nevertheless, looking at the bottom-right panel of Fig.~\ref{fig:M4}, the fraction of satellites is still high and
is in the range of $20$ to $30$ per cent. This is broadly consistent with the result of Model 2 shown in the right panel of 
Fig.~\ref{fig:M2}.
Thus the anisotropic clustering on large scales serves as a good probe of the fraction of satellites
robustly against different assumptions in the model.

\subsection{Robustness of the parameters against cosmological uncertainties}
\label{subsec:robust}
So far we have seen that the fraction of satellites is an important ingredient to understand the anisotropic
clustering of LRGs in redshift space. The discussion so far is, however, done keeping the underlying cosmological
model fixed to the one adopted in the simulations. We relax the cosmological assumptions in this section, 
and discuss the possibility of a simultaneous determination with cosmological parameters.
Here we focus on the constraints on the parameters that describe 
the connection between LRGs and subhalos, and leave the cosmological implications to the next section.

\begin{figure}
\begin{center}
\includegraphics[height=7.4cm]{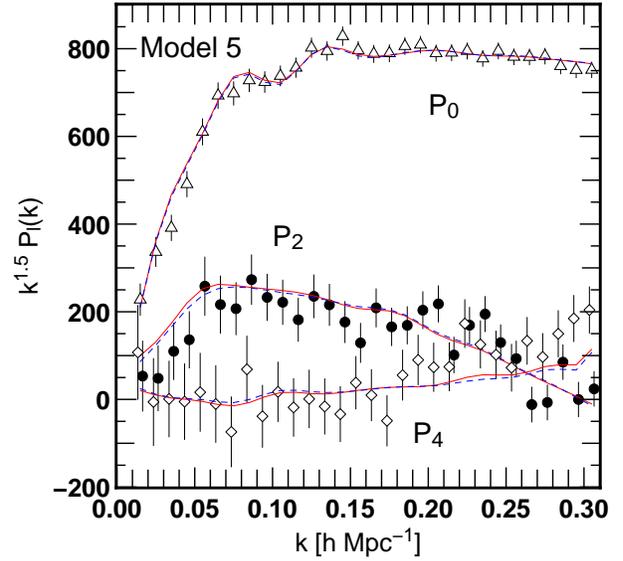}
\caption{Same as in Fig.~\ref{fig:P1}, but for Model 5.}
\label{fig:P5}
\end{center}
\end{figure}

We first show in Fig.~\ref{fig:P5} the best-fit power spectrum of Models 5a and 5b. Compared to the 
results of Model 2, the goodness of fit is improved ($\chi^2_\mathrm{red} = 1.18 \to 0.93$ for Model a and
$1.03 \to 0.95$ for Model b). Looking at each of the three multipoles, we can see that most of the improvement comes 
from the monopole moment (see Table~\ref{tab:best}).
We discuss in more detail on this improvement in the subsequent subsection.

\begin{figure}
\begin{center}
\includegraphics[height=3.8cm]{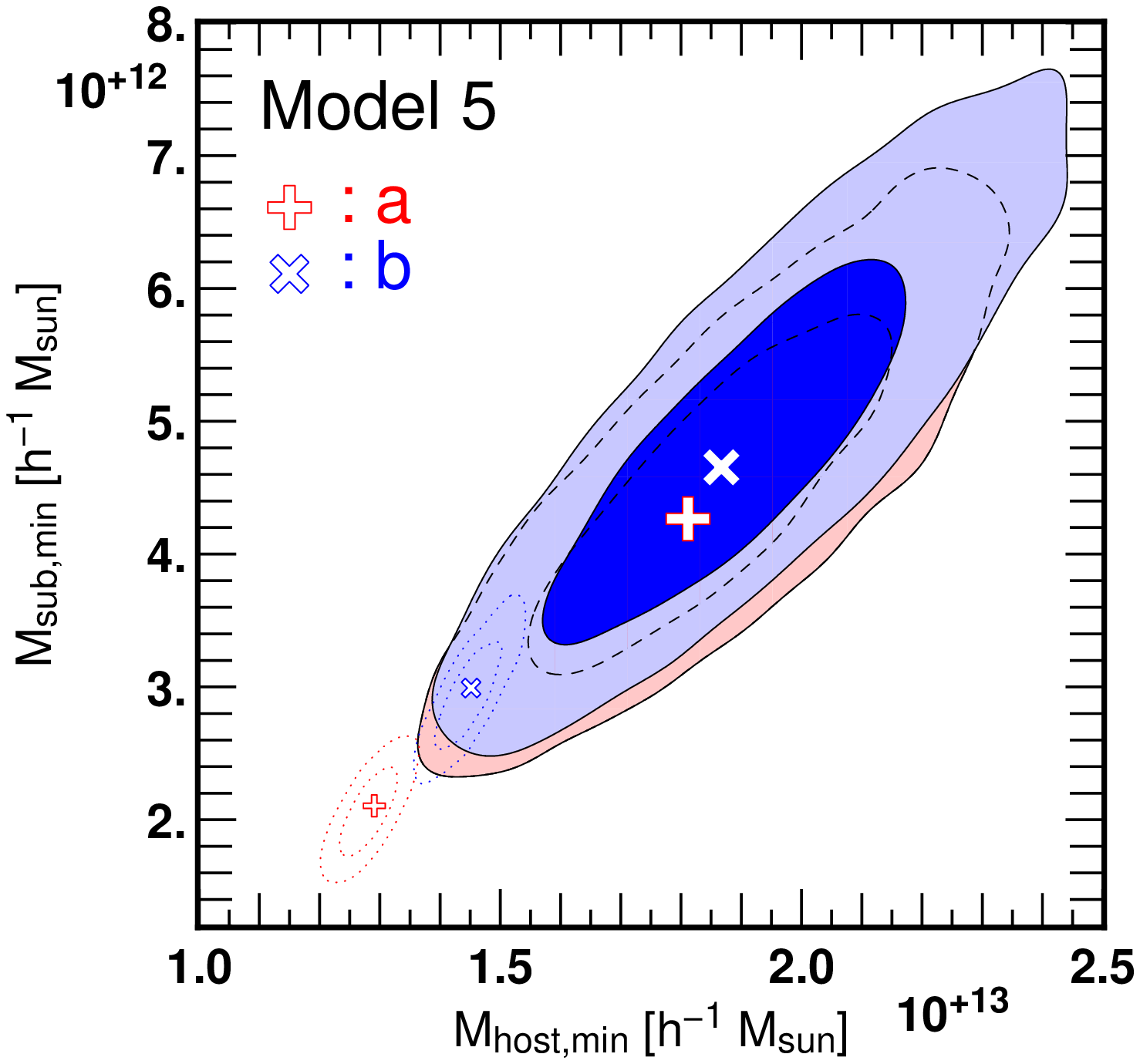}
\includegraphics[height=3.8cm]{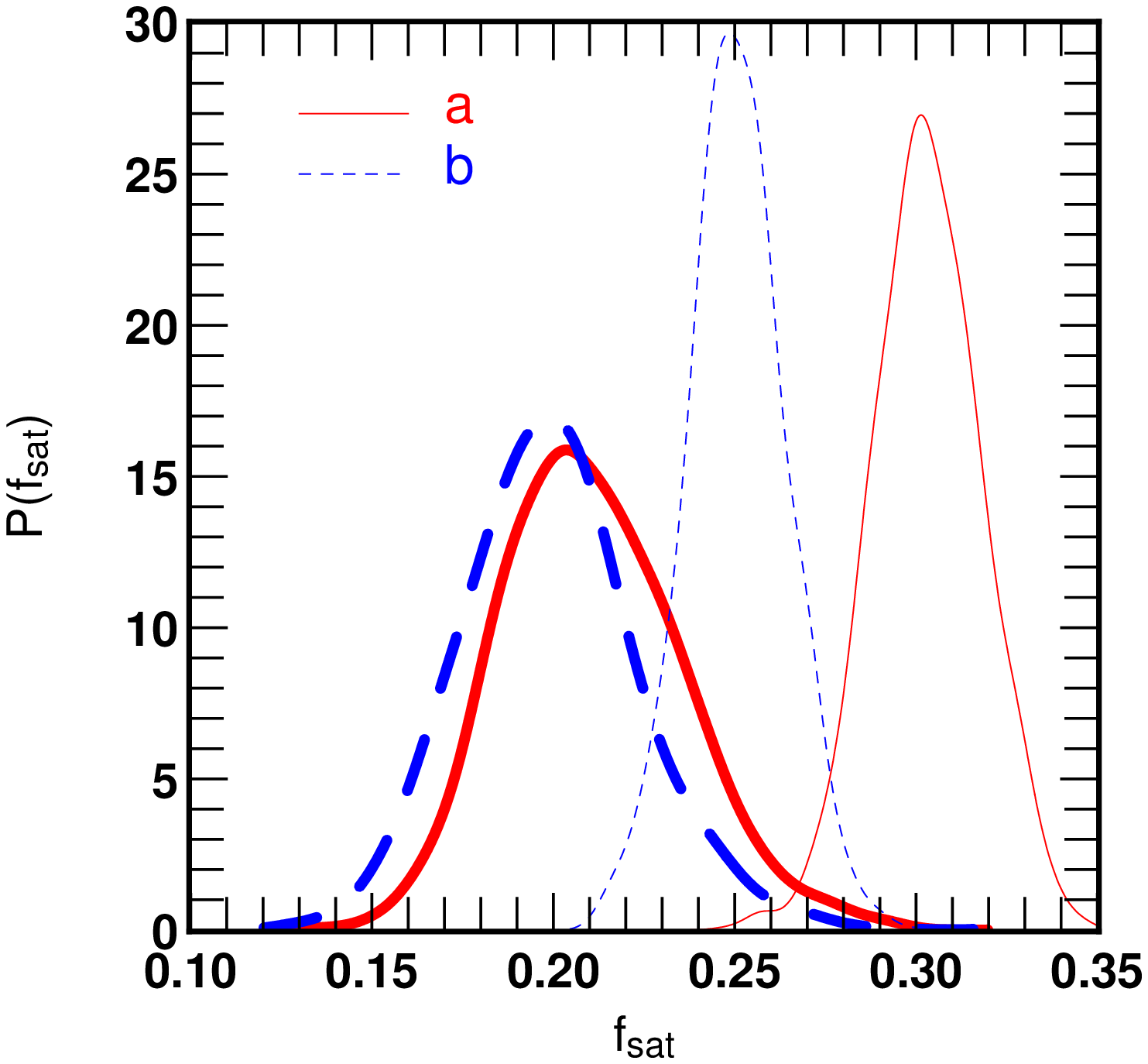}
\caption{Constraints on the parameters describing the properties of LRGs when cosmological parameters 
are varied (Model 5). We plot the $1$ and $2\sigma$ allowed regions for $M_\mathrm{host,min}$ and $M_\mathrm{sub,min}$ 
(colored contours in the left panel) and the fraction of satellites (thick lines in the right panel). 
We also plot the results when cosmological parameters are fixed (Model 2) 
by dotted contour lines in the left panel and thin lines in the right panel.}
\label{fig:M5}
\end{center}
\end{figure}

The constraints on the parameters, $M_\mathrm{host, min}$ and $M_\mathrm{sub, min}$ 
are shown in the left panel of Fig.~\ref{fig:M5}. We also show the
result of Model 2 in dotted contour lines. Compared to Model 2, the results are greatly altered in two ways: 
first, the best-fit minimum masses are shifted towards larger values and secondly the statistical uncertainties 
on them are increased by a factor of about five.

We plot in the right panel of Fig.~\ref{fig:M5} the satellite fraction derived with Model 5 (thick lines) 
and with Model 2 (thin lines).
Compared with the result in the left panel, the statistical uncertainty on the fraction of satellites 
does not increase so dramatically when cosmological assumptions are relaxed, although a smaller fraction is favored 
for Model 5 ($\sim 20$ per cent).
This basically confirms that the fraction of satellites still plays a dominant role to explain the anisotropy of the apparent
clustering pattern of LRGs even when we introduce additional sources of anisotropy.

Another interesting difference compared to Model 2 is that the difference between Model 5a and 5b are
much smaller than that between Model 2a and 2b. 
As we will discuss shortly, the distortion parameters introduced in Model 5 explains the observed anisotropy of
the clustering of LRGs partly, and the difference between Model 5a and 5b are somewhat absorbed by these new parameters.
Our final estimate of the $1\sigma$ ($2\sigma$) allowed region of the 
satellite fraction marginalized over cosmological uncertainties
is $f_\mathrm{sat}=0.214^{+0.018}_{-0.030}$($^{+0.046}_{-0.050}$) for Model 5a and 
$f_\mathrm{sat}=0.201^{+0.018}_{-0.028}$($^{+0.047}_{-0.050}$) for Model 5b.

\subsection{Prospects to derive cosmological parameters using $N$-body simulations}
\label{subsec:cosmo}
\begin{figure}
\begin{center}
\includegraphics[height=7.4cm]{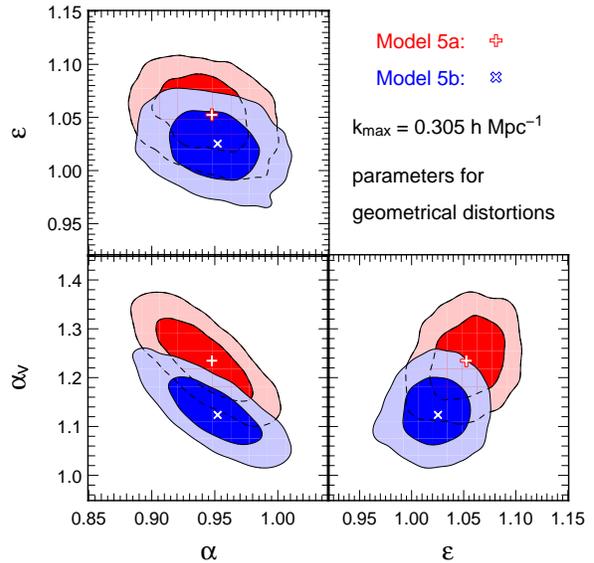}
\caption{Constraints on the distortion parameters marginalized over $M_\mathrm{host,min}$ and $M_\mathrm{sub,min}$.
We plot the $1$ and $2\sigma$ allowed regions of the parameters, $\alpha=(\alpha_\parallel\alpha_\perp^2)^{1/3}$,
$\epsilon=\alpha_\parallel/\alpha_\perp$, and $\alpha_\mathrm{v}$. We include wavenumbers up to 
$k_\mathrm{max}=0.305h\mathrm{Mpc}^{-1}$ into the analysis.}
\label{fig:geom}
\end{center}
\end{figure}

Now we turn to the cosmological constraints marginalized over the parameters that describe the nature of LRGs.
We first show the constraints on our parameters that induce distortions to the clustering of mock LRGs 
in Fig.~\ref{fig:geom}.
We plot the $67$ and $95$ per cent confidence regions of $\alpha_\mathrm{v}$ and two parameter combinations, 
$\alpha\equiv(\alpha_\parallel\alpha_\perp^2)^{1/3}$ and $\epsilon\equiv\alpha_\parallel/\alpha_\perp$.
The parameter $\alpha$ is responsible for the mismatch between the true and assumed distance scale, and
is expected to be determined by the BAO feature in the monopole moment. On the other hand, 
the other parameter, $\epsilon$, determines the significance of the AP effect that induces a geometrical distortion, 
and thus is sensitive to the higher multipoles.

Figure~\ref{fig:geom} suggests that the different modeling of LRGs using subhalos (i.e., Model 5a and 5b)
leads to different values of the distortion parameters favored by the observation.
We can find that the parameter $\alpha$ is the least affected among the three by the difference in
the prescription for mock LRGs. This is presumably owing to the robustness of the distance determination using the
feature of BAOs against different bias models. In both models, $\alpha<1$ is strongly suggested, and the best-fit
parameter is around $0.95$.
As we discussed in Section~\ref{subsec:RSD}, the sound horizon scale, $r_\mathrm{s}$, in our
fiducial cosmology is larger than that recently suggested. Our constraint on $\alpha$ is consistent with this
expectation, and supports recent observational results.
A value of $\alpha$ smaller than unity helps to reduce $\chi^2_{0,\mathrm{min}}$ from that in Model 2, 
and we can actually observe an improved fit to the BAO wiggles in the monopole moment in Fig.~\ref{fig:P5}.
Note also that $\chi^2_{0,\mathrm{min}}$ is the smallest in Model 5b (and almost the same in 5a)
compared to more complicated descriptions for the LRG-subhalo correspondence in Models 3 or 4.

The situation seems a bit different for the other two parameters.
They are more sensitive to the detail of the model and the confidence regions for the two models, 5a and 5b, 
have some non-negligible offset in the bottom-right panel of Fig.~\ref{fig:geom}.
In both cases, $\alpha_\mathrm{v}>1$ and $\epsilon>1$ are suggested.
In particular, Model 5a needs a large velocity boost factor, $\alpha_\mathrm{v}\sim1.2$.
Since the mock LRGs in this model have a smaller velocity dispersion than in Model 5b, 
a larger value of $\alpha_\mathrm{v}$ is required to explain the observed power spectrum.
This situation is quite similar to the results of Model 2a and 2b: we have shown that we need a larger satellite fraction
for Model 2a than Model 2b (see Figure~\ref{fig:M2}). Instead of the fraction of satellite, the parameter $\alpha_\mathrm{v}$
mainly absorbs the difference between the two models in this case.

Since both $\alpha_\mathrm{v}$ and $\epsilon$ are related to the amplitude of the apparent anisotropy in redshift space,
it is natural to see that they are degenerate with each other.
While a value of $\alpha_\mathrm{v}$ larger than unity {\it squashes} the apparent clustering pattern on large scales and 
{\it elongates} it on small scales along the line-of-sight direction 
(the Kaiser and the Fingers-of-God effect, respectively), a large $\epsilon$ always
{\it elongates} the apparent clustering independent of the distance scale. 
As we will show later, the large velocity boost factor, $\alpha_\mathrm{v}$, 
mainly comes from the power spectrum on small scales (i.e., a prominent Fingers-of-God suppression 
in the observed spectrum).
Then, a large $\alpha_\mathrm{v}$ determined on small scales might result in a overprediction of the Kaiser distortion 
on large scales. A large $\epsilon$ is chosen such that it partly cancels the strong Kaiser effect. 
This way, the difference in the modeling of velocities in the two models propagates to the derived 
distortion parameter, $\epsilon$.

It is worth noting that both models statistically exclude the fiducial parameter set, 
$\alpha_\mathrm{v}=\alpha=\epsilon=1$. This suggests that 
the cosmological model assumed in the simulations and in the redshift-distance relation might be different from the underlying
true cosmology:
the BAO scale in the simulation is too long, mock LRGs need larger velocities, and a geometrical distortion is induced
in the observed clustering to elongate the structure along the line of sight.
Though it might be too early to falsify the cosmological model from this analysis alone,
our results are encouraging. We can tell the difference between
our model based on simulations and the clustering of LRGs in the real universe with the current accuracy of the
observational data. As we will outline in Section~\ref{sec:discussion}, it might be very interesting to repeat the analysis
with some additional cosmological simulations and a new measurement of the power spectrum
assuming a cosmological model suggested by the current analysis.

\begin{figure}
\begin{center}
\includegraphics[height=7.4cm]{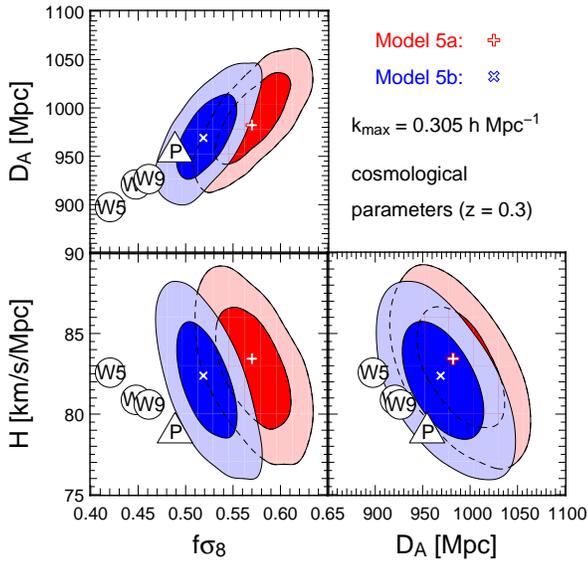}
\caption{Cosmological constraints. We assume the value of $r_\mathrm{s}$ 
from the PLANCK observation when we convert the constraints on $\alpha_\parallel$ and
$\alpha_\perp$ to $H$ and $D_\mathrm{A}$. We also show by symbols the results of CMB experiments 
(W5, W7, W9 and P respectively refer to WMAP5, WMAP7, WMAP9 and PLANCK; best-fit flat $\Lambda$CDM 
to the CMB data alone).}
\label{fig:cosmo_raw}
\end{center}
\end{figure}

We then derive the constraints on the parameters, $f\sigma_8$, 
$H$ and $D_\mathrm{A}$ using the relations, equations~(\ref{eq:cosmo1}), (\ref{eq:cosmo2}) and (\ref{eq:cosmo3}).
In doing so, we assume the value of $r_\mathrm{s}$ derived from the PLANCK observation (see Table~\ref{tab:cosmo}).
The results are shown in Fig.~\ref{fig:cosmo_raw}. For reference, we also mark the best-fit $\Lambda$CDM models
from some CMB observations by symbols (W5, W7 and W9 for WMAP, and P for PLANCK in the figure, and see
Table~\ref{tab:cosmo} for detail).
Our constraints in the $H$-$D_\mathrm{A}$ plane (bottom right) are broadly consistent with 
the CMB observations (interestingly, more consistent to more recent results).
On the other hand, the derived growth-rate parameter, $f\sigma_8$, is larger than the predictions of the
cosmological models favored by CMB observations.
We should interpret this with extreme caution.
The discrepancy between the Model 5a and 5b in the parameter combination, $f\sigma_8$, is as large as
$10$ per cent, and thus our estimate might have a systematic error of this size.
As we will see shortly, the constraint on $f\sigma_8$ from this analysis is likely to be contaminated significantly by nonlinearity
of the velocity field, and the derived combination might not be a linear growth rate but its nonlinear counterpart.

\begin{figure}
\begin{center}
\includegraphics[height=7.4cm]{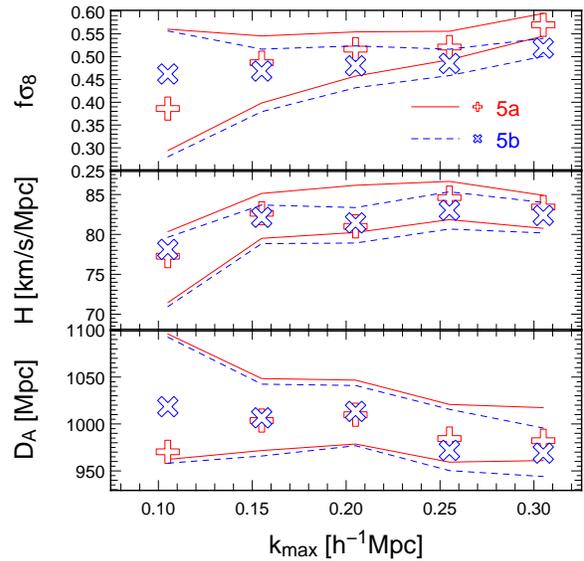}
\caption{Robustness of the results against the maximum wavenumber.
The $67$ per cent confidence intervals of 
the derived cosmological parameters are plotted as a function of the maximum wavenumber
(bands in solid lines for Model 5a, dashed lines for 5b).
The best-fit parameters are shown in plus symbols for Model 5a and in cross symbols for 5b.}
\label{fig:kmax}
\end{center}
\end{figure}

For a deeper understanding of the situation, 
we change the wavenumber range to be taken into account in the analysis, and discuss the sensitivity
of the results against the maximum wavenumber, $k_\mathrm{max}$.
The resulting one-dimensional marginalized constraints on the three cosmological parameters are 
shown in Fig.~\ref{fig:kmax}.
We show the 1$\sigma$ confidence intervals in the bands (solid for Model 5a, dashed for 5b).
When we adopt a value of $k_\mathrm{max}$ as small as $\sim 0.1\,h\,\mathrm{Mpc}^{-1}$,
the intervals for the two models almost coincide with each other. 
This suggests that the difference in the velocity structure on small scales is not relevant for the
anisotropy on very large scales.
The discrepancy between the two
models grows with $k_\mathrm{max}$, and finally the two $1\sigma$ intervals become exclusive for $f\sigma_8$ 
when $k_\mathrm{max}=0.305\,h\,\mathrm{Mpc}^{-1}$, which is our fiducial value. 

The parameter $f\sigma_8$ appears to be monotonically increasing with $k_\mathrm{max}$, and Model 5a 
shows a more sensitive response to $k_\mathrm{max}$ than 5b.
As we have already discussed, our estimate of $f\sigma_8$ through $\alpha_\mathrm{v}$ is affected by nonlinearity of the 
velocity field, and indeed the result indicates a scale-dependent velocity bias between our
mock LRGs and the observation. Larger velocities are required on smaller scales. 
Our estimator of $f\sigma_8$ is expected to have no dependence on $k_\mathrm{max}$ when the simulations
are perfect (i.e., $\alpha_\mathrm{v} = 1$ independent of $k_\mathrm{max}$). 
Since the amplitude of the velocity perturbations is expect to be linear at the large scale limit and our two
models give almost the same answer at $k_\mathrm{max}=0.105\,h\,\mathrm{Mpc}^{-1}$, the result at 
$k_\mathrm{max}=0.105\,h\,\mathrm{Mpc}^{-1}$ can be considered as a weak but reliable constraint on the linear growth rate.
We will comment on a possible solution for the nonlinear contaminations to the estimation of $f\sigma_8$ later in 
Section~\ref{sec:discussion}, and leave further investigations of the reliability of the constraint to future studies.

The other two parameters, $H$ and $D_\mathrm{A}$, show a weaker dependence on $k_\mathrm{max}$.
Although the situation seems to be better for these parameters than $f\sigma_8$, 
one should still take the results with caution.
The incompleteness of our modeling of the velocities can propagate to these parameters.
The parameters, $H$ and $D_\mathrm{A}$, are determined through $\alpha$ and $\epsilon$.
We show that $\alpha$ is robust against systematics in Fig.~\ref{fig:geom}.
Nevertheless, a misestimation of $f\sigma_8$ propagates to $\epsilon$, that results in a systematic error on $H$
and $D_\mathrm{A}$.
It is of interest to further test the constraints on these parameters with new simulations that can reproduce the observed
RSDs without a large velocity boost $\alpha_\mathrm{v}$, and a study along this line is now ongoing.

\begin{figure}
\begin{center}
\includegraphics[height=7.4cm]{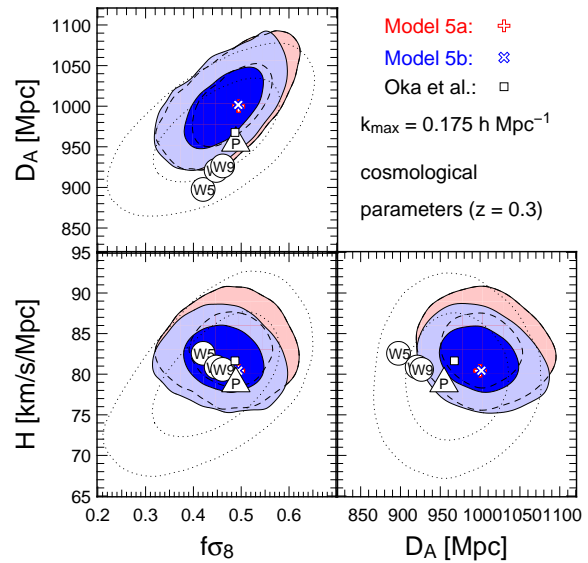}
\caption{Comparison between our cosmological constraints (colored contours) and those obtained by 
fitting the same data but with an analytical model described in the accompanying paper, (\citealt{Oka13}; dotted contours). 
We change the maximum wavenumber to $k_\mathrm{max}=0.175\,h\,\mathrm{Mpc}^{-1}$ so as to
match with that in \citet{Oka13}.}
\label{fig:cosmo_raw2}
\end{center}
\end{figure}

We have discussed so far the constraints on the cosmological parameters using $N$-body simulations
and their possible systematics. Now we compare our results with those obtained with an independent method.
We plot in Fig.~\ref{fig:cosmo_raw2} the constraints on the three cosmological parameters from our analysis
(colored contours) and those by the accompanying paper \citep{Oka13} in which we adopt an analytical model 
for the parameter estimation (contours in dashed lines with the best-fit values depicted by cubes).
In this figure, we adopt $k_\mathrm{max}=0.175\,h\,\mathrm{Mpc}^{-1}$, which is the same as in \citet{Oka13}.

Overall, the contours obtained by \citet{Oka13} enclose the allowed regions obtained by our analysis.
Especially, the best-fit parameters obtained with different models
are consistent with each other at about 1$\sigma$ level.
This is encouraging since the two analyses rely on totally different prescriptions for the nonlinear growth of structure,
the redshift-space distortions and the galaxy bias.
The weaker constraints by \citet{Oka13} probably come from the fact that the analytical model has 
free parameters to control the
galaxy bias as well as the velocity dispersion in the analytical model. These nuisance parameters are effectively
fixed in our analysis once the connection between LRGs and subhalos is assumed.

The result suggests that it is potentially possible to tighten the constraints by adopting a reasonable prescription
for galaxies using subhalos identified in simulations. Alternatively, giving a prior information to the velocity dispersion 
based on simulations rather than letting it float as a free parameter might be another way to obtain tighter constraints.
The best-fit parameters derived from CMB observations are in good agreement with the contours obtained by 
\citet{Oka13}. If we believe the contours obtained in this study, we can exclude some of the CMB results.
This suggests that the anisotropic clustering of galaxies have a strong statistical power if one successfully
models the nonlinear growth of the velocity field (velocity dispersion, in this particular case) and the galaxy bias.

\subsection{Multiplicity function}
\label{subsec:multi}
So far we have focused on the clustering of LRGs in redshift space 
on relatively large scales and discussed how well we can explain the observational data with 
mock LRGs in cosmological simulations.
It is of course important to construct a mock galaxy catalog that has a mean number density consistent with observations.
Indeed mock galaxy catalog is usually constructed based on HOD or abundance matching techniques,
in which {\it one-point} statistics are directly drawn from observation or constrained from 
clustering information on {\it small scales}. In these methods, mock galaxies are modeled so that the 
observed mean number density is recovered automatically.
It is interesting to check the compatibility of our mock LRGs with observed one point statistics.
In this section, we compare our best-fit models with the multiplicity function estimated by \citet{Reid09}.

Before making a comparison, we note some differences in the LRG samples used in our study and 
in \citet{Reid09}.
As summarized in Section~\ref{sec:data}, our sample is based on the DR7 and we use LRGs only on the north cap
in the redshift range of $0.16 < z < 0.47$. By contrast, the sample analyzed in \citet{Reid09} is based on DR4+
\citep{SDSSDR5} and they include both caps but restrict the redshift range to $0.16 < z < 0.36$.
They also supplementary add the imaging sample to correct for fiber collisions, 
incomplete sky and complex angular masks. 
It is thus not straightforward to directly compare our results with theirs, but nevertheless the comparison
is still meaningful to test our alternative mock making scheme and infer the nature of LRGs.

First of all, the mean number density of mock LRGs in some of our best-fit models are significantly larger than
the observation. The sample in \citet{Reid09} has $n_\mathrm{g}\simeq9.7\times 10^{-5}\,h^3\,\mathrm{Mpc}^{-3}$
\citep{Zehavi05}, and this is two to four times smaller than the results of Models 2, 3 and 5 (see Table~\ref{tab:best}). 
On the other hand, Model 4 has a slightly smaller number density than that from observation. 

This discrepancy in the mean number density implies that the mass of the halos as well as that of subhalos do not
uniquely determine whether a LRG can live there or not: our model simply assumes that {\it all} the subhalos above a 
mass threshold have LRGs, and this assumption is so simple that lead the number density larger than the observed.
A simple way towards a more sophisticated modeling of LRGs is to introduce a second physical parameter 
that determines the habitability of a LRG.
An example is the presence of a massive halo at an earlier epoch (i.e., $z=2$ or $3$) adopted in the
abundance matching technique \citep{Masaki13}.
This approach was shown to recover the halo occupation distribution and the angular clustering very well
in addition to the number density, which is automatically adjusted by the matching by construction.

Instead of pursuing along this direction, we adopt a simple statistical approach
to adjust the mean density in this study.
We conduct random sampling and 
reduce the number of mock LRGs assuming that the observed LRG catalog is a Poisson sample of our mock LRGs.
As we have already noted, this procedure does not affect the expectation value of the power spectrum
only resulting in a larger shot noise and thus
the successful fitting results in the previous sections are not diminished.
Also, the fraction of satellites derived by our analysis is not affected by random sampling.

We might be able to justify the adjustment of the number density with random sampling as follows.
Galaxy bias is a 
{\it stochastic} process by nature and is difficult to connect simulated subhalos to observed LRGs {\it deterministically}
\citep{Dekel99}.
In other words, there might exist some unknown parameters to determine whether a subhalo hosts a LRG  
or not \citep{Taruya00}. 
Modeling LRGs from the first principle is well beyond the scope of this study based on $N$-body simulations 
that solves a purely gravitational system, and baryonic physics must be essential in order to fully predict 
(i.e., deterministically simulate) the clustering of LRGs or galaxies in general (e.g., \citealt{Somerville01,Yoshikawa01}).
We give up a deterministic prediction of LRGs but instead effectively take into account the stochasticity of galaxy bias
by random sampling. 
In what follows, we adjust the spatial density to $n_\mathrm{g} = 10^{-4}\,h^3\,\mathrm{Mpc}^{-3}$ for 
Models 2, 3 and 5, while we keep all the mock LRGs in Model 4.

\begin{figure}
\begin{center}
\includegraphics[height=7.4cm]{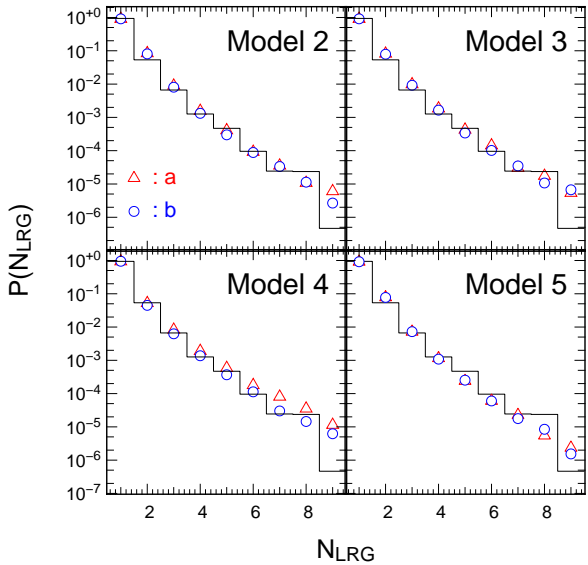}
\caption{Multiplicity function of LRGs from our best-fit models (symbols) against observation (histogram).
The triangles (circles) show the result of Model a (b).}
\label{fig:multi1}
\end{center}
\end{figure}

We now compare the multiplicity function with that measured by \citet{Reid09}. We denote the number of LRGs in
a host halo by $N_\mathrm{LRG}$. In our case, the host halos are simply the FoF groups while
\citet{Reid09} identify groups with the Count-in-Cylinder method. Then the multiplicity function, 
$P(N_\mathrm{LRG})$, is defined as the fraction of LRG systems with $N_\mathrm{LRG}$ members.
Note that we do not count FoF groups without a mock LRG in the denominator to match the definition of the observed 
multiplicity function.
We plot in Fig.~\ref{fig:multi1} $P(N_\mathrm{LRG})$ against $N_\mathrm{LRG}$.
The overall decreasing trend of $P(N_\mathrm{LRG})$ with $N_\mathrm{LRG}$ seen in the observed data (histograms) 
is well explained by all the models (symbols; triangles for Model a and circles for Model b). 
This coincidence is notable since we construct the catalogs guided only by the clustering pattern in redshift space 
on large scales.
Also, remember that the allowed regions of the parameters characterizing the condition of subhalos to host LRGs
are significantly different in different models.
These results suggest that the multiplicity function is an important ingredient that determines the 
multipole moments of the power spectrum in redshift space on large scales.

\begin{figure}
\begin{center}
\includegraphics[height=7.4cm]{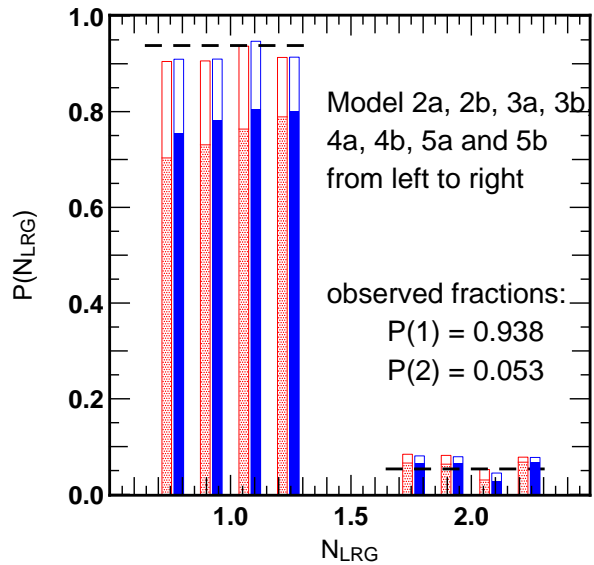}
\caption{Zoom of Fig.~\ref{fig:multi1} for $N_\mathrm{LRG} = 1$ and $2$. The filled part of the bars
show the contribution for systems with a central LRG and the open part is responsible for the systems without
a central LRG.}
\label{fig:multi2}
\end{center}
\end{figure}

We plot in Fig.~\ref{fig:multi2} the first two bins, $N_\mathrm{LRG} = 1$ and $2$ in linear scale.
This time the models are shown in bars (Model 2a, 2b, ..., 5b from left to right), 
while the observational data are shown by horizontal dashed lines.
The filled part of the bars shows mock LRG systems that contain a central LRG, 
while the open part indicates those consisting of only satellites.
The observational data as well as our models suggest that most of the LRGs are 
distributed in a single- or double-LRG systems, and these two bins mostly determine 
the clustering properties. The fraction of single-LRG systems ranges from $90$ to $95$ per cent with our eight models,
and this number is in good agreement with the observed fraction, $93.8$ per cent. Model 4a and 4b give
the most consistent fractions to the observation, and might be a better description of LRGs than other models.
Indeed, these models have the smallest $\chi^2_\mathrm{red}$ to the multipole moments (see Table~\ref{tab:best}).

An important difference in the interpretation of the observed multiplicity function in our study and in \citet{Reid09}
is the definition of centrals and satellites.
Our definition of central LRGs are those hosted by subhalos that have the largest mass in the same host halos,
and the rest of LRGs are satellites.
Consequently, the most massive subhalos sometimes do not host a LRG in our model and
a LRG system can be exclusively composed of satellite LRGs.
On the other hand, in \citet{Reid09}, only LRGs in multiple-LRG systems are considered as satellites and
all the single-LRG systems are regarded as central LRGs.
The satellite fraction as large as $20$ to $30$ per cent derived in our analysis mainly comes from this difference.
If we instead employ the same definition of satellites as in \citet{Reid09}, the fraction becomes much smaller.
We list in Table~\ref{tab:best} the fraction of satellites, $f_\mathrm{sat}^*$, derived with this definition.
The values are now $6$ to $10$ per cent, and are roughly the same as the observation. 
Thus the seemingly large satellite fraction of our mock LRGs does not conflict with the result of \citet{Reid09},
in which they derive a satellite fraction of $6.36^{+0.38}_{-0.39}$.

One can say that our definition is based on the kinematics of LRGs: what matters to explain the feature of RSDs in 
observation is how many LRGs have velocities relative to their host halos.
One important lesson we have learned through this analysis is that a non-negligible fraction of 
LRGs without a close companion (i.e., single-LRG systems) are not located at the center of the host halos. 
The situation is consistent with the result of a cross correlation analysis of single-LRG systems 
with galaxy-galaxy lensing signal 
(and also cross correlation with the photometric galaxies) performed by \citet{Hikage13a}.
Their result suggests that about $24$ per cent of these LRGs have offset from the true gravitational center of the systems
(see Figure~10 in the reference).

\subsection{Halo occupation distribution}
\label{subsec:HOD}
The halo occupation distribution (HOD) is a useful approach to connect galaxies to dark matter halos frequently 
adopted in the literature. This function is closely related to the multiplicity function presented in the previous subsection.
Indeed, \citet{Reid09} gives a tight constraint on the HOD by fitting the observed multiplicity function.
We here compare the HOD of our models with that in the literature and discuss the compatibility of a model based
on the observed HOD to explain the anisotropic clustering of LRGs.

\begin{figure}
\begin{center}
\includegraphics[width=8.4cm]{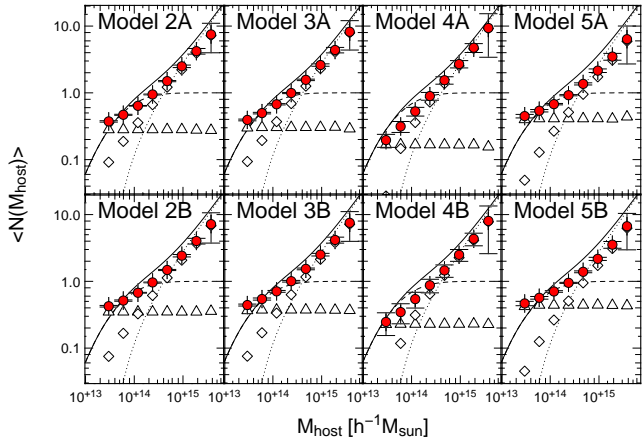}
\caption{Mean halo occupation number as a function of the halo mass derived from our eight models.
Triangles and diamonds show the central and satellite contributions, respectively, while the sum
is plotted as circles with error bars. Also shown by lines is the model described by equation~(\ref{eq:HODtot}) 
with the parameters determined by \citet{Reid09}. 
Solid, dashed, and dotted lines show the total, central, and satellite HOD, respectively.}
\label{fig:HOD_ourmodels}
\end{center}
\end{figure}

Figure~\ref{fig:HOD_ourmodels} shows the HODs measured from our mock catalogs with the best-fit parameters.
Centrals and satellites are respectively plotted by triangles and diamonds, while the sum of the two populations
are shown as circles with error bar showing the scatter among different random realizations.
Also shown by lines are the functions,
\be
N_\mathrm{tot} &=& N_\mathrm{cen}(1 + N_\mathrm{sat}),\label{eq:HODtot}\\
N_\mathrm{cen} &=& \frac{1}{2}\left[1+\mathrm{erf}\left(\frac{\log_{10}M_\mathrm{host}-
\log_{10}M_\mathrm{min}}{\sigma}\right)\right],\label{eq:HODcen}\\
N_\mathrm{sat} &=& \left(\frac{M_\mathrm{host}-M_\mathrm{cut}}{M_1}\right)^\alpha,\label{eq:HODsat}
\ee
with the parameters, $M_\mathrm{min} = 5.796\times10^{13} h^{-1}M_\odot$, 
$M_\mathrm{cut}=3.6\times10^{13} h^{-1}M_\odot$, 
$M_1=3.564\times10^{14} h^{-1}M_\odot$, $\sigma=0.7$ and $\alpha=1.035$ determined by \citet{Reid09} based 
on the observation.
The dashed and dotted lines are the central and satellite contribution and the sum is depicted by the solid line.

Overall, the total HOD is very similar among our eight models, and its slope is close to that by \citet{Reid09}.
There exists, however, an offset in the overall amplitude.
This is likely to be due to the difference in the halo mass function between ours and that in \citet{Reid09}:
the halo finder (FOF and SO), output redshifts (z=0.35 and 0.2), and a different cosmological model 
(WMAP 5yrs and 3yrs) can give a noticeable change in the halo mass function.
With a different mass function, the HOD that gives the correct mean number density should be different.
Another difference in the HOD can be seen at the low mass end.
It is natural for our models to have a rather sharp cut because we impose a minimum
halo mass to host a LRG. We need a more elaborate model to have a smoother HOD.

When focusing on each of the central and satellite contributions, the difference among the models are clearer.
The central contribution is smaller for models in which we randomly discard more central subhalos
(i.e., Model 4), while
the observed HOD is unity when the host halo mass is large. As already discussed in the previous subsection,
an important point is that it is the HOD of the sum of the two populations that is directly constrained from observation,
and the contributions from each population strongly depends on the definition of central and satellite LRGs.
We thus do not consider the different HOD for each population problematic.

\begin{figure}
\begin{center}
\includegraphics[width=8.4cm]{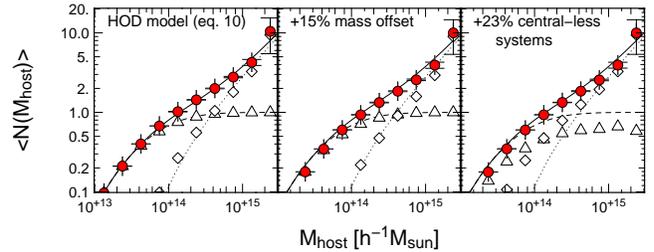}
\caption{Mean halo occupation number as a function of the halo mass for models based on the HOD by
\citet{Reid09} (lines). We follow the central HOD in equation~(\ref{eq:HODcen}) and then assign satellites
to subhalos more massive than $10^{13}\,h^{-1}M_\odot$ whose host halos have a central LRG to
obtain almost identical HOD in the left panel. 
We scale the masses by $15\%$ to adjust the mean number density in the middle panel. We finally
replace centrals with satellites randomly following the procedure described in the text in the right panel.}
\label{fig:HOD_models}
\end{center}
\end{figure}

Given the similar but different HODs, we next examine whether the HOD derived from observation
can explain the anisotropic clustering on large scales at the same time. We do this by employing again
the subhalos found in the simulations, but enforcing that the HOD is the same as that in \citet{Reid09}.
We here consider the center-of-mass positions and velocities for simplicity.
We first assign LRGs to central subhalos with the probability of equation~(\ref{eq:HODcen}) depending on the mass of the
host halos.
We then consider halos with a central LRG and assign satellite LRGs to other member
subhalos. We empirically find that we can recover the HOD for satellites given by equation (\ref{eq:HODsat}) almost perfectly
when we assign satellite LRGs to subhalos with mass larger than $10^{13}\,h^{-1}M_\odot$.
The resultant HOD is shown in the left panel of Fig.~\ref{fig:HOD_models}.
Despite almost the same HOD, the mean number density of these mock LRGs is about $20\%$ larger
than the observed one. This is again because of the difference in the halo mass function.
We adjust the mean number density by scaling the masses, 
$M_\mathrm{min}$, $M_\mathrm{cut}$ and $M_1$ in the HOD function by $15\%$ (middle panel).

\begin{figure}
\begin{center}
\includegraphics[width=8.4cm]{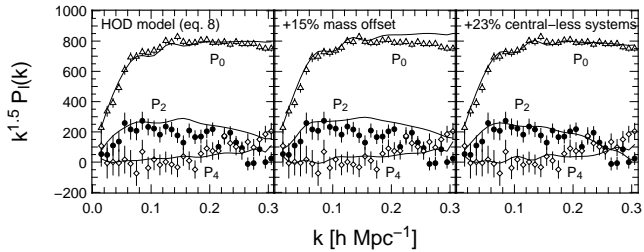}
\caption{Multipole moments of the power spectrum measured from the mock LRGs constructed based on the observed HOD.
The three panels corresponds to those in Fig.~\ref{fig:HOD_models}.}
\label{fig:power_HOD}
\end{center}
\end{figure}

We compare the multipole moments of the power spectrum in redshift space measured from these new
mock LRGs in the left and middle panels of Fig.~\ref{fig:power_HOD} before and after the mass scaling, respectively.
We can see in the left panel that the amplitude and the slope of the monopole and quadrupole moments are 
somewhat different from observed moments. Indeed the chi-squared statistics for these mocks is $222.1$ for $90$ 
data points, which is significantly larger than what we find for our other mock catalogs discussed so far.
After the mass scaling, the amplitude of the monopole moment is now closer to the observation on large scales 
($k\simlt0.15\,h\,\mathrm{Mpc}^{-1}$) at the cost of larger gap on smaller scales as well as on the quadrupole moment
(middle panel). The value of chi-squared gets larger, $405.1$, after shifting the masses.

The middle panel of Fig.~\ref{fig:power_HOD} suggests that the observed multipole moments are affected 
more strongly by the Fingers-of-God suppression than the mock LRGs constructed based on the observed HOD.
Motivated by the high satellite fraction derived from the main analysis, we consider the following simple model.
Starting from the mock catalogs with the HOD shown in the middle panel of Fig.~\ref{fig:HOD_models}, 
we randomly select a central LRG and reassign it to the most massive satellite subhalo without a LRG in the
same host halo.
Note that the total number of LRGs within a host halo is conserved throughout this process.
We repeat this procedure to $23\%$ of the halos with a central LRG to finally have a satellite fraction of 
$30\%$ without changing the total HOD (see right panel of Figure~\ref{fig:HOD_models}). 
We now have some LRG systems, which have only satellites. 
Note that such systems do not exist in the other two mock catalogs discussed in this subsection, nor in the 
model described by equation (\ref{eq:HODtot}) by construction.
We finally show the power spectrum measured from the mock LRGs with increased satellite fraction in the right panel
of Fig.~\ref{fig:power_HOD}. We now have an excellent agreement with observation with the chi-squared statistics
being $82.0$. This is comparable to the other models discussed in previous subsections.

We conclude that the large satellite fraction is always necessary to explain the multipole moments of the power spectrum
independent of the detail of the model. This subsection also demonstrates that one needs a statistical procedure such
as random sampling to reduce the total number or random swapping of centrals to satellites to explain 
both the mean number density (and also the HOD) and the clustering in redshift space, as long as we
rely only on the mass of halos and of subhalos.
Although more involved models with further physical inputs such as the formation epoch of the subhalos \citep{Masaki13} 
would be an interesting next step, we leave it to a future investigation given the already successful fit to the observation.

\section{Summary and discussions}
\label{sec:discussion}
The clustering statistics of galaxies on large scales provides us with a wealth of cosmological information.
A simple parametric approach is usually adopted to absorb the uncertainty of galaxy bias.
This simple treatment is still expected to work well
as long as one restricts the analysis to sufficiently large scales
where the nature of galaxies does not affect the statistics significantly.
We consider in this study a bias model for LRGs based on subhalos identified in a series of 
cosmological $N$-body simulations to pursue more realistic bias model for analyses on large scales.
We determine the parameters that characterize the condition of subhalos to host a LRG by directly fitting the 
first three non-zero multipole moments of the power spectrum in redshift space on scales $k\simlt0.3\,h\,\mathrm{Mpc}^{-1}$. 
We find that models employing both central and satellite subhalos successfully 
reproduce the clustering pattern within the statistical error, while a model with the central population alone clearly fails to do so. 

An important ingredient in the model is the fraction of the satellite LRGs, and this parameter
is constrained to $20$ to $30$ per cent slightly depending on the detail of the model.
Though the large scale clustering alone cannot constrain the detail of the model such as the HOD,
the fraction of satellites is shown to be robust against the assumptions in the model.
Indeed, we cannot explain the multipole moments unless we increase the satellite fraction by hand to $30$ per cent
even when we adopt the observed HOD of LRGs.

Since our model for LRGs is rather simple relying only on the mass of the host halos and of subhalos,
it is difficult to explain the clustering and number density simultaneously without a statistical procedures
such as random sampling or the random conversion of LRG hosts from centrals into satellites.
We definitely need a more involved model or eventually need proper account of baryonic physics to have a prediction 
without random processes. We expect that these complications are more responsible for the clustering 
properties on small scales where the one-halo contribution is dominant. Although the range of wavenumber 
that we consider here is naively expected to be in a regime where the impact of nonlinear physics is mild,
it would be important to test the compatibility of these models against observation based on a proper statistical argument.

Although we use the simulation output obtained for one particular cosmological model, 
we induce additional anisotropic clustering signal by deforming the simulation box and changing the magnitude of 
redshift-space distortions. By doing this, we demonstrate that we can determine cosmological and LRG parameters
simultaneously. The constraints on the distances ($H$ and $D_\mathrm{A}$) are found to be more robust against different 
prescriptions for LRGs thanks to the clear signal of baryon acoustic oscillations in the observed
spectrum, while nonlinear corrections can give a systematic correction to the measurement of the growth-rate parameter 
$f\sigma_8$, when the modeling of the velocities of LRGs is imperfect.
We here propose a possible extension of the current analysis and a solution to the nonlinear contamination to
the measurement of the growth rate. 

Our constraints on the parameters $\alpha_\mathrm{v}$, $\alpha_\parallel$ and $\alpha_\perp$ suggest
that the cosmological model adopted in our simulations is slightly disfavored by the observation.
The most significant deviation from the fiducial value is seen in the parameter $\alpha_\mathrm{v}$ especially
on small scales.
A simple step to go beyond the present analysis is to re-simulate the clustering of LRGs with the best-fit
cosmological model obtained here and repeat the same analysis again.
Also, we suggest to repeat the measurement of the power spectrum from the observation assuming a new
distance-redshift relation. 
When the cosmological model adopted in the simulations as well as in the distance-redshift relation 
is close enough to the true underlying model, 
these deformation parameters should be consistent with unity (i.e., no deformation, no velocity bias). 
We can iterate the analysis until this condition is satisfied.
Although it is computationally too expensive to simulate the clustering over a cosmological parameter space, 
several times of iterations are well within reach of the current power of super computers.
Since we wish to constrain the growth-rate parameter in the {\it linear} regime, it might be helpful to introduce
one more parameter in our model which controls the motion of mock LRGs relative to the host halos.
The growth-rate parameter less affected by nonlinearity may be derived after marginalizing over this new parameter
in return for a looser constraint in the fraction of satellites.

Before presenting the final estimate of the cosmological parameters with these modifications to the analysis, 
several issues must be carefully treated. One of our approximate treatment is in the estimate of the statistical error.
We ignore here the off-diagonal components of the covariance matrix of the observational data
and also the statistical error in the template power spectrum
measured from the simulations in finite volume. 
We discuss the current level of accuracy of the error estimation in Appendix~\ref{sec:quality}.
Another is the effect of fiber collisions which is ignored in measuring the power spectrum analyzed in this study.
Although this effect is smaller than the current statistical accuracy on large scales (see Appendix~\ref{sec:fiber}), 
it should be properly included in the construction of mock LRGs when we extend the analysis to smaller scales.
Finally, the result of the analysis may depend on the subhalo finder \citep{Pujol13}.
We naively expect that the dependence is most prominent in the parameters that describe the nature of galaxies
(e.g., the minimum halo mass to host a galaxy).
However, the constraints on the cosmological parameters should also be
verified carefully especially when the statistical error becomes very small with future ambitious survey projects.

We believe that this study is an important first step toward cosmological analyses using simulations as a theoretical template.
The use of simulations allows to take into account the nonlinear and nonlocal nature of galaxy bias at full order.
Also, our analysis demonstrates that we can break the degeneracy
between the uncertainty in the nature of galaxies and the cosmological model. An important ingredient in this
analysis is the fraction of satellite LRGs, and we show that we can still put a meaningful constraint on this
parameter when we relax the assumption of the cosmological model.
This is encouraging to us to continue along this line and apply our methodology to some future galaxy surveys
targeting galaxies whose environmental properties are not understood very well.

\section*{Acknowledgment}
We appreciate Naoki Yoshida for kindly providing us a numerical code to identify subhalos, 
which is implemented in our analysis.
We acknowledge Kazuhiro Yamamoto for the observational data and valuable comments on the effect
of fiber collisions. We also thank Atsushi Taruya, Shun Saito and Yasushi Suto 
for fruitful discussions on the cosmological implications.
Numerical computations for the present work have been carried out on Cray XC30 at Center for Computational 
Astrophysics, CfCA, of National Astronomical Observatory of Japan.
TN is supported by Japan Society for the Promotion of Science (JSPS) Postdoctoral Fellowships for Research Abroad. 

\footnotesize{
\bibliographystyle{mn2e}
\bibliography{lssref} 

\begin{thebibliography}{}
 \providecommand{\href}[2]{#2}
  \providecommand{\doi}[1]{\href{http://dx.doi.org/#1}{doi:#1}}
  \providecommand{\eprint}[1]{\href{http://arxiv.org/abs/#1}{arXiv:#1}}

\bibitem[\protect\citeauthoryear{{Abazajian} et~al.,}{{Abazajian}
  et~al.}{2009}]{SDSSDR7}
{Abazajian} K.~N.  et~al., 2009, ApJS, 182, 543

\bibitem[\protect\citeauthoryear{{Adelman-McCarthy} et~al.,}{{Adelman-McCarthy}
  et~al.}{2007}]{SDSSDR5}
{Adelman-McCarthy} J.~K.  et~al., 2007, ApJS, 172, 634

\bibitem[\protect\citeauthoryear{{Alcock} \& {Paczynski}}{{Alcock} \&
  {Paczynski}}{1979}]{Alcock79}
{Alcock} C.,  {Paczynski} B.,  1979, \nat, 281, 358

\bibitem[\protect\citeauthoryear{{Anderson} et~al.,}{{Anderson}
  et~al.}{2014}]{Anderson:2014kq}
{Anderson} L.  et~al., 2014, \mnras, 439, 83

\bibitem[\protect\citeauthoryear{{Angulo} \& {White}}{{Angulo} \&
  {White}}{2010}]{Angulo10}
{Angulo} R.~E.,  {White} S.~D.~M.,  2010, \mnras, 405, 143

\bibitem[\protect\citeauthoryear{{Anselmi}, {Matarrese} \&
  {Pietroni}}{{Anselmi} et~al.}{2011}]{Anselmi11}
{Anselmi} S.,  {Matarrese} S.,    {Pietroni} M.,  2011, JCAP, 6, 15

\bibitem[\protect\citeauthoryear{{Ballinger}, {Peacock} \&
  {Heavens}}{{Ballinger} et~al.}{1996}]{Ballinger96}
{Ballinger} W.~E.,  {Peacock} J.~A.,    {Heavens} A.~F.,  1996, \mnras, 282,
  877

\bibitem[\protect\citeauthoryear{{Berlind} \& {Weinberg}}{{Berlind} \&
  {Weinberg}}{2002}]{Berlind:2002kx}
{Berlind} A.~A.,  {Weinberg} D.~H.,  2002, \apj, 575, 587

\bibitem[\protect\citeauthoryear{{Bernardeau}, {Crocce} \&
  {Scoccimarro}}{{Bernardeau} et~al.}{2008a}]{Bernardeau08}
{Bernardeau} F.,  {Crocce} M.,    {Scoccimarro} R.,  2008a, \prd, 78, 103521

\bibitem[\protect\citeauthoryear{{Bernardeau}, {Van de Rijt} \&
  {Vernizzi}}{{Bernardeau} et~al.}{2013b}]{Bernardeau13}
{Bernardeau} F.,  {Van de Rijt} N.,    {Vernizzi} F.,  2013b, \prd, 87, 043530

\bibitem[\protect\citeauthoryear{{Bernardeau}, {Colombi}, {Gazta{\~n}aga} \&
  {Scoccimarro}}{{Bernardeau} et~al.}{2002c}]{Bernardeau02}
{Bernardeau} F.,  {Colombi} S.,  {Gazta{\~n}aga} E.,    {Scoccimarro} R.,
  2002c, Phys. Rep., 367, 1

\bibitem[\protect\citeauthoryear{{Beutler} et~al.,}{{Beutler}
  et~al.}{2012}]{Beutler12}
{Beutler} F.  et~al., 2012, \mnras, 423, 3430

\bibitem[\protect\citeauthoryear{{Blake}, {Collister} \& {Lahav}}{{Blake}
  et~al.}{2008a}]{Blake08}
{Blake} C.,  {Collister} A.,    {Lahav} O.,  2008a, \mnras, 385, 1257

\bibitem[\protect\citeauthoryear{{Blake} et~al.,}{{Blake}
  et~al.}{2011b}]{Blake11a}
{Blake} C.  et~al., 2011b, \mnras, 415, 2876

\bibitem[\protect\citeauthoryear{{Blake} et~al.,}{{Blake}
  et~al.}{2011c}]{Blake11c}
{Blake} C.  et~al., 2011c, \mnras, 418, 1707

\bibitem[\protect\citeauthoryear{{Blake} et~al.,}{{Blake}
  et~al.}{2012d}]{Blake12}
{Blake} C.  et~al., 2012d, \mnras, 425, 405

\bibitem[\protect\citeauthoryear{{Carlson}, {White} \& {Padmanabhan}}{{Carlson}
  et~al.}{2009}]{Carlson09}
{Carlson} J.,  {White} M.,    {Padmanabhan} N.,  2009, \prd, 80, 043531

\bibitem[\protect\citeauthoryear{{Chuang} \& {Wang}}{{Chuang} \&
  {Wang}}{2012a}]{Chuang12}
{Chuang} C.-H.,  {Wang} Y.,  2012a, \mnras, 426, 226

\bibitem[\protect\citeauthoryear{{Chuang} \& {Wang}}{{Chuang} \&
  {Wang}}{2013b}]{Chuang13a}
{Chuang} C.-H.,  {Wang} Y.,  2013b, \mnras, 431, 2634

\bibitem[\protect\citeauthoryear{{Chuang} \& {Wang}}{{Chuang} \&
  {Wang}}{2013c}]{Chuang13b}
{Chuang} C.-H.,  {Wang} Y.,  2013c, \mnras, 435, 255

\bibitem[\protect\citeauthoryear{{Cole}}{{Cole}}{1997}]{Cole97}
{Cole} S.,  1997, \mnras, 286, 38

\bibitem[\protect\citeauthoryear{{Conroy}, {Wechsler} \& {Kravtsov}}{{Conroy}
  et~al.}{2006}]{Conroy06}
{Conroy} C.,  {Wechsler} R.~H.,    {Kravtsov} A.~V.,  2006, \apj, 647, 201

\bibitem[\protect\citeauthoryear{{Crocce} \& {Scoccimarro}}{{Crocce} \&
  {Scoccimarro}}{2006a}]{Crocce06b}
{Crocce} M.,  {Scoccimarro} R.,  2006a, \prd, 73, 063519

\bibitem[\protect\citeauthoryear{{Crocce} \& {Scoccimarro}}{{Crocce} \&
  {Scoccimarro}}{2006b}]{Crocce06c}
{Crocce} M.,  {Scoccimarro} R.,  2006b, \prd, 73, 063520

\bibitem[\protect\citeauthoryear{{Crocce} \& {Scoccimarro}}{{Crocce} \&
  {Scoccimarro}}{2008c}]{Crocce08}
{Crocce} M.,  {Scoccimarro} R.,  2008c, \prd, 77, 023533

\bibitem[\protect\citeauthoryear{{Crocce}, {Pueblas} \& {Scoccimarro}}{{Crocce}
  et~al.}{2006a}]{Crocce06a}
{Crocce} M.,  {Pueblas} S.,    {Scoccimarro} R.,  2006a, \mnras, 373, 369

\bibitem[\protect\citeauthoryear{{Crocce}, {Scoccimarro} \&
  {Bernardeau}}{{Crocce} et~al.}{2012b}]{Crocce12}
{Crocce} M.,  {Scoccimarro} R.,    {Bernardeau} F.,  2012b, \mnras, 427, 2537

\bibitem[\protect\citeauthoryear{{Davis}, {Efstathiou}, {Frenk} \&
  {White}}{{Davis} et~al.}{1985}]{Davis:1985fk}
{Davis} M.,  {Efstathiou} G.,  {Frenk} C.~S.,    {White} S.~D.~M.,  1985, \apj,
  292, 371

\bibitem[\protect\citeauthoryear{{Dekel} \& {Lahav}}{{Dekel} \&
  {Lahav}}{1999}]{Dekel99}
{Dekel} A.,  {Lahav} O.,  1999, \apj, 520, 24

\bibitem[\protect\citeauthoryear{{Eisenstein} \& {Hu}}{{Eisenstein} \&
  {Hu}}{1998}]{Eisenstein98}
{Eisenstein} D.~J.,  {Hu} W.,  1998, Astrophys. J., 496, 605

\bibitem[\protect\citeauthoryear{{Eisenstein} et~al.,}{{Eisenstein}
  et~al.}{2001a}]{Eisenstein01}
{Eisenstein} D.~J.  et~al., 2001a, AJ, 122, 2267

\bibitem[\protect\citeauthoryear{{Eisenstein} et~al.,}{{Eisenstein}
  et~al.}{2005b}]{Eisenstein05}
{Eisenstein} D.~J.  et~al., 2005b, \apj, 633, 560

\bibitem[\protect\citeauthoryear{{Eisenstein}, {Seo} \& {White}}{{Eisenstein}
  et~al.}{2007c}]{Eisenstein07}
{Eisenstein} D.~J.,  {Seo} H.-J.,    {White} M.,  2007c, \apj, 664, 660

\bibitem[\protect\citeauthoryear{{Feldman}, {Kaiser} \& {Peacock}}{{Feldman}
  et~al.}{1994}]{Feldman94}
{Feldman} H.~A.,  {Kaiser} N.,    {Peacock} J.~A.,  1994, Astrophys. J., 426,
  23

\bibitem[\protect\citeauthoryear{{Guo}, {White}, {Angulo}, {Henriques},
  {Lemson}, {Boylan-Kolchin}, {Thomas} \& {Short}}{{Guo} et~al.}{2013a}]{Guo13}
{Guo} Q.,  {White} S.,  {Angulo} R.~E.,  {Henriques} B.,  {Lemson} G.,
  {Boylan-Kolchin} M.,  {Thomas} P.,    {Short} C.,  2013a, \mnras, 428, 1351

\bibitem[\protect\citeauthoryear{{Guo}, {Zehavi} \& {Zheng}}{{Guo}
  et~al.}{2012b}]{Guo12}
{Guo} H.,  {Zehavi} I.,    {Zheng} Z.,  2012b, \apj, 756, 127

\bibitem[\protect\citeauthoryear{{Guzzo} et~al.,}{{Guzzo}
  et~al.}{2008}]{Guzzo08}
{Guzzo} L.  et~al., 2008, \nat, 451, 541

\bibitem[\protect\citeauthoryear{{Hikage} \& {Yamamoto}}{{Hikage} \&
  {Yamamoto}}{2013}]{Hikage13b}
{Hikage} C.,  {Yamamoto} K.,  2013, JCAP, 8, 19

\bibitem[\protect\citeauthoryear{{Hikage}, {Mandelbaum}, {Takada} \&
  {Spergel}}{{Hikage} et~al.}{2013}]{Hikage13a}
{Hikage} C.,  {Mandelbaum} R.,  {Takada} M.,    {Spergel} D.~N.,  2013, \mnras,
  435, 2345

\bibitem[\protect\citeauthoryear{{Hinshaw} et~al.,}{{Hinshaw}
  et~al.}{2013}]{WMAP9}
{Hinshaw} G.  et~al., 2013, ApJS, 208, 19

\bibitem[\protect\citeauthoryear{{Hiramatsu} \& {Taruya}}{{Hiramatsu} \&
  {Taruya}}{2009}]{Hiramatsu09}
{Hiramatsu} T.,  {Taruya} A.,  2009, \prd, 79, 103526

\bibitem[\protect\citeauthoryear{{Hockney} \& {Eastwood}}{{Hockney} \&
  {Eastwood}}{1981}]{Hockney81}
{Hockney} R.~W.,  {Eastwood} J.~W.,  1981, {Computer Simulation Using
  Particles}

\bibitem[\protect\citeauthoryear{{Ho}, {Lin}, {Spergel} \& {Hirata}}{{Ho}
  et~al.}{2009}]{Ho09}
{Ho} S.,  {Lin} Y.-T.,  {Spergel} D.,    {Hirata} C.~M.,  2009, \apj, 697,
  1358

\bibitem[\protect\citeauthoryear{{Hu} \& {Haiman}}{{Hu} \&
  {Haiman}}{2003}]{Hu03}
{Hu} W.,  {Haiman} Z.,  2003, \prd, 68, 063004

\bibitem[\protect\citeauthoryear{{Hu} \& {Sawicki}}{{Hu} \&
  {Sawicki}}{2007}]{Hu:2007lr}
{Hu} W.,  {Sawicki} I.,  2007, \prd, 76, 064004

\bibitem[\protect\citeauthoryear{{Jackson}}{{Jackson}}{1972}]{Jackson72}
{Jackson} J.~C.,  1972, \mnras, 156, 1P

\bibitem[\protect\citeauthoryear{{Jeong} \& {Komatsu}}{{Jeong} \&
  {Komatsu}}{2006a}]{Jeong06}
{Jeong} D.,  {Komatsu} E.,  2006a, \apj, 651, 619

\bibitem[\protect\citeauthoryear{{Jeong} \& {Komatsu}}{{Jeong} \&
  {Komatsu}}{2009b}]{Jeong09b}
{Jeong} D.,  {Komatsu} E.,  2009b, \apj, 691, 569

\bibitem[\protect\citeauthoryear{{Jing}}{{Jing}}{2005}]{Jing05}
{Jing} Y.~P.,  2005, \apj, 620, 559

\bibitem[\protect\citeauthoryear{{Kaiser}}{{Kaiser}}{1987a}]{Kaiser87}
{Kaiser} N.,  1987a, \mnras, 227, 1

\bibitem[\protect\citeauthoryear{{Kaiser}}{{Kaiser}}{1984b}]{Kaiser84}
{Kaiser} N.,  1984b, Astrophys. J. L., 284, L9

\bibitem[\protect\citeauthoryear{{Kazin} et~al.,}{{Kazin}
  et~al.}{2013}]{Kazin13}
{Kazin} E.~A.  et~al., 2013, \mnras, 435, 64

\bibitem[\protect\citeauthoryear{{Komatsu} et~al.,}{{Komatsu}
  et~al.}{2009a}]{WMAP5}
{Komatsu} E.  et~al., 2009a, Astrophys. J. Suppl., 180, 330

\bibitem[\protect\citeauthoryear{{Komatsu} et~al.,}{{Komatsu}
  et~al.}{2011b}]{WMAP7}
{Komatsu} E.  et~al., 2011b, Astrophys. J. S., 192, 18

\bibitem[\protect\citeauthoryear{{Kulkarni}, {Nichol}, {Sheth}, {Seo},
  {Eisenstein} \& {Gray}}{{Kulkarni} et~al.}{2007}]{Kulkarni07}
{Kulkarni} G.~V.,  {Nichol} R.~C.,  {Sheth} R.~K.,  {Seo} H.-J.,  {Eisenstein}
  D.~J.,    {Gray} A.,  2007, \mnras, 378, 1196

\bibitem[\protect\citeauthoryear{{Lewis}, {Challinor} \& {Lasenby}}{{Lewis}
  et~al.}{2000}]{CAMB}
{Lewis} A.,  {Challinor} A.,    {Lasenby} A.,  2000, Astrophys. J., 538, 473

\bibitem[\protect\citeauthoryear{{Linder}}{{Linder}}{2008}]{Linder08}
{Linder} E.~V.,  2008, Astroparticle Physics, 29, 336

\bibitem[\protect\citeauthoryear{{Mandelbaum}, {Seljak}, {Kauffmann}, {Hirata}
  \& {Brinkmann}}{{Mandelbaum} et~al.}{2006}]{Mandelbaum06}
{Mandelbaum} R.,  {Seljak} U.,  {Kauffmann} G.,  {Hirata} C.~M.,    {Brinkmann}
  J.,  2006, \mnras, 368, 715

\bibitem[\protect\citeauthoryear{{Masaki}, {Hikage}, {Takada}, {Spergel} \&
  {Sugiyama}}{{Masaki} et~al.}{2013}]{Masaki13}
{Masaki} S.,  {Hikage} C.,  {Takada} M.,  {Spergel} D.~N.,    {Sugiyama} N.,
  2013, \mnras, 433, 3506

\bibitem[\protect\citeauthoryear{{Matarrese} \& {Pietroni}}{{Matarrese} \&
  {Pietroni}}{2007a}]{Matarrese07}
{Matarrese} S.,  {Pietroni} M.,  2007a, JCAP, 6, 26

\bibitem[\protect\citeauthoryear{{Matarrese} \& {Pietroni}}{{Matarrese} \&
  {Pietroni}}{2008b}]{Matarrese08a}
{Matarrese} S.,  {Pietroni} M.,  2008b, Modern Physics Letters A, 23, 25

\bibitem[\protect\citeauthoryear{{Matsubara}}{{Matsubara}}{2008a}]{Matsubara08%
a}
{Matsubara} T.,  2008a, \prd, 77, 063530

\bibitem[\protect\citeauthoryear{{Matsubara}}{{Matsubara}}{2008b}]{Matsubara08%
b}
{Matsubara} T.,  2008b, \prd, 78, 083519

\bibitem[\protect\citeauthoryear{{Matsubara} \& {Suto}}{{Matsubara} \&
  {Suto}}{1996}]{Matsubara:1996qy}
{Matsubara} T.,  {Suto} Y.,  1996, \apjl, 470, L1

\bibitem[\protect\citeauthoryear{{Ma} \& {Fry}}{{Ma} \&
  {Fry}}{2000}]{Ma:2000lr}
{Ma} C.-P.,  {Fry} J.~N.,  2000, \apj, 543, 503

\bibitem[\protect\citeauthoryear{{McCullagh}, {Neyrinck}, {Szapudi} \&
  {Szalay}}{{McCullagh} et~al.}{2013}]{McCullagh13}
{McCullagh} N.,  {Neyrinck} M.~C.,  {Szapudi} I.,    {Szalay} A.~S.,  2013,
  ApJL, 763, L14

\bibitem[\protect\citeauthoryear{{McDonald}}{{McDonald}}{2006a}]{McDonald06}
{McDonald} P.,  2006a, \prd, 74, 103512

\bibitem[\protect\citeauthoryear{{McDonald}}{{McDonald}}{2007b}]{McDonald07}
{McDonald} P.,  2007b, \prd, 75, 043514

\bibitem[\protect\citeauthoryear{{Mead} \& {Peacock}}{{Mead} \&
  {Peacock}}{2013}]{Mead13}
{Mead} A.,  {Peacock} J.,  2014, \mnras, 440, 1233

\bibitem[\protect\citeauthoryear{{Nishimichi} \& {Taruya}}{{Nishimichi} \&
  {Taruya}}{2011}]{Nishimichi11}
{Nishimichi} T.,  {Taruya} A.,  2011, \prd, 84, 043526

\bibitem[\protect\citeauthoryear{{Nishimichi} et~al.,}{{Nishimichi}
  et~al.}{2007a}]{Nishimichi07}
{Nishimichi} T.  et~al., 2007a, PASJ, 59, 1049

\bibitem[\protect\citeauthoryear{{Nishimichi} et~al.,}{{Nishimichi}
  et~al.}{2009b}]{Nishimichi09}
{Nishimichi} T.  et~al., 2009b, Publ. Astron. Soc. Japan, 61, 321

\bibitem[\protect\citeauthoryear{{Nishizawa}, {Takada} \&
  {Nishimichi}}{{Nishizawa} et~al.}{2013}]{Nishizawa13}
{Nishizawa} A.~J.,  {Takada} M.,    {Nishimichi} T.,  2013, \mnras, 433, 209

\bibitem[\protect\citeauthoryear{{Okamura}, {Taruya} \& {Matsubara}}{{Okamura}
  et~al.}{2011}]{Okamura11}
{Okamura} T.,  {Taruya} A.,    {Matsubara} T.,  2011, JCAP, 8, 12

\bibitem[\protect\citeauthoryear{{Oka}, {Saito}, {Nishimichi}, {Taruya} \&
  {Yamamoto}}{{Oka} et~al.}{2014}]{Oka13}
{Oka} A.,  {Saito} S.,  {Nishimichi} T.,  {Taruya} A.,    {Yamamoto} K.,  2014,\mnras, 439, 2515

\bibitem[\protect\citeauthoryear{{Okumura}, {Matsubara}, {Eisenstein}, {Kayo},
  {Hikage}, {Szalay} \& {Schneider}}{{Okumura} et~al.}{2008}]{Okumura08}
{Okumura} T.,  {Matsubara} T.,  {Eisenstein} D.~J.,  {Kayo} I.,  {Hikage} C.,
  {Szalay} A.~S.,    {Schneider} D.~P.,  2008, \apj, 676, 889

\bibitem[\protect\citeauthoryear{{Padmanabhan} \& {White}}{{Padmanabhan} \&
  {White}}{2008a}]{Padmanabhan08}
{Padmanabhan} N.,  {White} M.,  2008a, \prd, 77, 123540

\bibitem[\protect\citeauthoryear{{Padmanabhan} \& {White}}{{Padmanabhan} \&
  {White}}{2009b}]{Padmanabhan09b}
{Padmanabhan} N.,  {White} M.,  2009b, \prd, 80, 063508

\bibitem[\protect\citeauthoryear{{Padmanabhan}, {White}, {Norberg} \&
  {Porciani}}{{Padmanabhan} et~al.}{2009}]{Padmanabhan09a}
{Padmanabhan} N.,  {White} M.,  {Norberg} P.,    {Porciani} C.,  2009, \mnras,
  397, 1862

\bibitem[\protect\citeauthoryear{{Peacock} \& {Smith}}{{Peacock} \&
  {Smith}}{2000}]{Peacock:2000qy}
{Peacock} J.~A.,  {Smith} R.~E.,  2000, \mnras, 318, 1144

\bibitem[\protect\citeauthoryear{{Percival} et~al.,}{{Percival}
  et~al.}{2004}]{Percival04}
{Percival} W.~J.  et~al., 2004, \mnras, 353, 1201

\bibitem[\protect\citeauthoryear{{Pietroni}}{{Pietroni}}{2008}]{Pietroni08}
{Pietroni} M.,  2008, JCAP, 10, 36

\bibitem[\protect\citeauthoryear{{Planck Collaboration} et~al.,}{{Planck
  Collaboration} et~al.}{2013}]{Planck_cosmo}
{Planck Collaboration} et~al., 2013, ArXiv e-prints, \eprint{1303.5076}

\bibitem[\protect\citeauthoryear{{Pujol} et~al.,}{{Pujol}
  et~al.}{2013}]{Pujol13}
{Pujol} A.  et~al., 2014, \mnras, 438, 3205

\bibitem[\protect\citeauthoryear{{Reid} \& {Spergel}}{{Reid} \&
  {Spergel}}{2009}]{Reid09}
{Reid} B.~A.,  {Spergel} D.~N.,  2009, \apj, 698, 143

\bibitem[\protect\citeauthoryear{{Reid} et~al.,}{{Reid} et~al.}{2012}]{Reid12}
{Reid} B.~A.  et~al., 2012, \mnras, 426, 2719

\bibitem[\protect\citeauthoryear{{Samushia} et~al.,}{{Samushia}
  et~al.}{2013}]{Samushia13}
{Samushia} L.  et~al., 2013, \mnras, 429, 1514

\bibitem[\protect\citeauthoryear{{Sato} \& {Matsubara}}{{Sato} \&
  {Matsubara}}{2011}]{Sato11}
{Sato} M.,  {Matsubara} T.,  2011, \prd, 84, 043501

\bibitem[\protect\citeauthoryear{{Scoccimarro}}{{Scoccimarro}}{1998}]{Scoccima%
rro98}
{Scoccimarro} R.,  1998, \mnras, 299, 1097

\bibitem[\protect\citeauthoryear{{Scoccimarro}, {Sheth}, {Hui} \&
  {Jain}}{{Scoccimarro} et~al.}{2001}]{Scoccimarro:2001fj}
{Scoccimarro} R.,  {Sheth} R.~K.,  {Hui} L.,    {Jain} B.,  2001, \apj, 546,
  20

\bibitem[\protect\citeauthoryear{{Seljak}}{{Seljak}}{2000}]{Seljak:2000uq}
{Seljak} U.,  2000, \mnras, 318, 203

\bibitem[\protect\citeauthoryear{{Sherwin} \& {Zaldarriaga}}{{Sherwin} \&
  {Zaldarriaga}}{2012}]{Sherwin12}
{Sherwin} B.~D.,  {Zaldarriaga} M.,  2012, \prd, 85, 103523

\bibitem[\protect\citeauthoryear{{Shoji}, {Jeong} \& {Komatsu}}{{Shoji}
  et~al.}{2009}]{Shoji09}
{Shoji} M.,  {Jeong} D.,    {Komatsu} E.,  2009, \apj, 693, 1404

\bibitem[\protect\citeauthoryear{{Simha} \& {Cole}}{{Simha} \&
  {Cole}}{2013}]{Simha13}
{Simha} V.,  {Cole} S.,  2013, \mnras, 436, 1142

\bibitem[\protect\citeauthoryear{{Somerville}, {Lemson}, {Sigad}, {Dekel},
  {Kauffmann} \& {White}}{{Somerville} et~al.}{2001}]{Somerville01}
{Somerville} R.~S.,  {Lemson} G.,  {Sigad} Y.,  {Dekel} A.,  {Kauffmann} G.,
  {White} S.~D.~M.,  2001, \mnras, 320, 289

\bibitem[\protect\citeauthoryear{{Song} \& {Percival}}{{Song} \&
  {Percival}}{2009}]{Song09}
{Song} Y.-S.,  {Percival} W.~J.,  2009, JCAP, 10, 4

\bibitem[\protect\citeauthoryear{{Springel}}{{Springel}}{2005}]{GADGET2}
{Springel} V.,  2005, \mnras, 364, 1105

\bibitem[\protect\citeauthoryear{{Springel}, {White}, {Tormen} \&
  {Kauffmann}}{{Springel} et~al.}{2001}]{SUBFIND}
{Springel} V.,  {White} S.~D.~M.,  {Tormen} G.,    {Kauffmann} G.,  2001,
  \mnras, 328, 726

\bibitem[\protect\citeauthoryear{{Starobinsky}}{{Starobinsky}}{2007}]{Starobin%
sky:2007fk}
{Starobinsky} A.~A.,  2007, Soviet Journal of Experimental and Theoretical
  Physics Letters, 86, 157

\bibitem[\protect\citeauthoryear{{Sugiyama} \& {Futamase}}{{Sugiyama} \&
  {Futamase}}{2012a}]{Sugiyama12a}
{Sugiyama} N.~S.,  {Futamase} T.,  2012a, ArXiv e-prints, \eprint{1210.7499}

\bibitem[\protect\citeauthoryear{{Sugiyama} \& {Futamase}}{{Sugiyama} \&
  {Futamase}}{2012b}]{Sugiyama12b}
{Sugiyama} N.~S.,  {Futamase} T.,  2012b, \apj, 760, 114

\bibitem[\protect\citeauthoryear{{Sugiyama} \& {Spergel}}{{Sugiyama} \&
  {Spergel}}{2013}]{Sugiyama13}
{Sugiyama} N.~S.,  {Spergel} D.~N.,  2014, JCAP, 2, 42

\bibitem[\protect\citeauthoryear{{S{\'a}nchez} et~al.,}{{S{\'a}nchez}
  et~al.}{2013}]{Sanchez13}
{S{\'a}nchez} A.~G.  et~al., 2013, \mnras, 433, 1202

\bibitem[\protect\citeauthoryear{{Takahashi} et~al.,}{{Takahashi}
  et~al.}{2009}]{Takahashi09}
{Takahashi} R.  et~al., 2009, \apj, 700, 479

\bibitem[\protect\citeauthoryear{{Taruya} \& {Hiramatsu}}{{Taruya} \&
  {Hiramatsu}}{2008}]{Taruya08}
{Taruya} A.,  {Hiramatsu} T.,  2008, \apj, 674, 617

\bibitem[\protect\citeauthoryear{{Taruya} \& {Suto}}{{Taruya} \&
  {Suto}}{2000}]{Taruya00}
{Taruya} A.,  {Suto} Y.,  2000, \apj, 542, 559

\bibitem[\protect\citeauthoryear{{Taruya}, {Nishimichi}, {Saito} \&
  {Hiramatsu}}{{Taruya} et~al.}{2009a}]{Taruya09}
{Taruya} A.,  {Nishimichi} T.,  {Saito} S.,    {Hiramatsu} T.,  2009a, \prd,
  80, 123503

\bibitem[\protect\citeauthoryear{{Taruya}, {Saito} \& {Nishimichi}}{{Taruya}
  et~al.}{2011b}]{Taruya11}
{Taruya} A.,  {Saito} S.,    {Nishimichi} T.,  2011b, \prd, 83, 103527

\bibitem[\protect\citeauthoryear{{Taruya}, {Bernardeau}, {Nishimichi} \&
  {Codis}}{{Taruya} et~al.}{2012c}]{Taruya12}
{Taruya} A.,  {Bernardeau} F.,  {Nishimichi} T.,    {Codis} S.,  2012c,
  \prd, 86, 103528

\bibitem[\protect\citeauthoryear{{Taruya}, {Nishimichi} \&
  {Bernardeau}}{{Taruya} et~al.}{2013d}]{Taruya13}
{Taruya} A.,  {Nishimichi} T.,    {Bernardeau} F.,  2013d, \prd, 87, 083509

\bibitem[\protect\citeauthoryear{{Tormen} \& {Bertschinger}}{{Tormen} \&
  {Bertschinger}}{1996}]{Tormen96}
{Tormen} G.,  {Bertschinger} E.,  1996, \apj, 472, 14

\bibitem[\protect\citeauthoryear{{Valageas}}{{Valageas}}{2013a}]{Valageas2013}
{Valageas} P.,  2013a, \prd, 88, 083524

\bibitem[\protect\citeauthoryear{{Valageas}}{{Valageas}}{2007b}]{Valageas07}
{Valageas} P.,  2007b, A\&A, 465, 725

\bibitem[\protect\citeauthoryear{{Valageas}}{{Valageas}}{2011c}]{Valageas11c}
{Valageas} P.,  2011c, A\&A, 526, A67

\bibitem[\protect\citeauthoryear{{Valageas} \& {Nishimichi}}{{Valageas} \&
  {Nishimichi}}{2011}]{Valageas11a}
{Valageas} P.,  {Nishimichi} T.,  2011, Astronomy \& Astrophysics, 527, A87

\bibitem[\protect\citeauthoryear{{Valageas}, {Nishimichi} \&
  {Taruya}}{{Valageas} et~al.}{2013}]{Valageas13a}
{Valageas} P.,  {Nishimichi} T.,    {Taruya} A.,  2013, \prd, 87, 083522

\bibitem[\protect\citeauthoryear{{Wake} et~al.,}{{Wake} et~al.}{2008}]{Wake08}
{Wake} D.~A.  et~al., 2008, \mnras, 387, 1045

\bibitem[\protect\citeauthoryear{{White}, {Zheng}, {Brown}, {Dey} \&
  {Jannuzi}}{{White} et~al.}{2007}]{White07}
{White} M.,  {Zheng} Z.,  {Brown} M.~J.~I.,  {Dey} A.,    {Jannuzi} B.~T.,
  2007, ApJL, 655, L69

\bibitem[\protect\citeauthoryear{{Xu}, {Cuesta}, {Padmanabhan}, {Eisenstein} \&
  {McBride}}{{Xu} et~al.}{2013}]{Xu13}
{Xu} X.,  {Cuesta} A.~J.,  {Padmanabhan} N.,  {Eisenstein} D.~J.,    {McBride}
  C.~K.,  2013, \mnras, 431, 2834

\bibitem[\protect\citeauthoryear{{Yamamoto}, {Nakamichi}, {Kamino}, {Bassett}
  \& {Nishioka}}{{Yamamoto} et~al.}{2006a}]{Yamamoto06}
{Yamamoto} K.,  {Nakamichi} M.,  {Kamino} A.,  {Bassett} B.~A.,    {Nishioka}
  H.,  2006a, PASJ, 58, 93

\bibitem[\protect\citeauthoryear{{Yamamoto}, {Nakamura}, {H{\"u}tsi},
  {Narikawa} \& {Sato}}{{Yamamoto} et~al.}{2010b}]{Yamamoto10}
{Yamamoto} K.,  {Nakamura} G.,  {H{\"u}tsi} G.,  {Narikawa} T.,    {Sato} T.,
  2010b, \prd, 81, 103517

\bibitem[\protect\citeauthoryear{{Yamamoto}, {Sato} \& {H{\"u}tsi}}{{Yamamoto}
  et~al.}{2008c}]{Yamamoto08}
{Yamamoto} K.,  {Sato} T.,    {H{\"u}tsi} G.,  2008c, Progress of Theoretical
  Physics, 120, 609

\bibitem[\protect\citeauthoryear{{York} et~al.,}{{York} et~al.}{2000}]{SDSS}
{York} D.~G.  et~al., 2000, AJ, 120, 1579

\bibitem[\protect\citeauthoryear{{Yoshikawa}, {Taruya}, {Jing} \&
  {Suto}}{{Yoshikawa} et~al.}{2001}]{Yoshikawa01}
{Yoshikawa} K.,  {Taruya} A.,  {Jing} Y.~P.,    {Suto} Y.,  2001, \apj, 558,
  520

\bibitem[\protect\citeauthoryear{{Zehavi} et~al.,}{{Zehavi}
  et~al.}{2005}]{Zehavi05}
{Zehavi} I.  et~al., 2005, \apj, 621, 22

\bibitem[\protect\citeauthoryear{{Zheng}, {Zehavi}, {Eisenstein}, {Weinberg} \&
  {Jing}}{{Zheng} et~al.}{2009}]{Zheng09}
{Zheng} Z.,  {Zehavi} I.,  {Eisenstein} D.~J.,  {Weinberg} D.~H.,    {Jing}
  Y.~P.,  2009, \apj, 707, 554

\end{thebibliography}
}

\appendix
\section{Calculation of the model multipoles and their statistical error}
\label{sec:quality}

In this Appendix, we show the accuracy of the multipole moments of the power spectrum measured from 
$N$-body simulations and discuss the validity of our approximate treatment of the covariance matrix.

First, we discuss the estimator of the multipoles used in their measurement from $N$-body simulations. 
One of a simple estimator adopts a binning scheme and assign each wavevector to a wavenumber bin:
\be
\hat{P}^{(n)}_{\ell,\mathrm{binned}}(k_i) = \frac{1}{N_{i}}\frac{2\ell+1}{2}
\sum_{\mathbf{k}\in i\mathrm{th\,bin}}\mathcal{P}_\ell(\mu_\mathbf{k})\left|\delta_\mathbf{k}^{(n)}\right|^2,
\label{eq:binned}
\ee
where $N_i$ stands for the number of modes available in the $i$th wavenumber bin.
This estimator, however, suffers from an artificial effect caused by the discreteness of wavevectors in finite volume.
On large scales, the number of modes $N_i$ is usually small, and therefore we cannot sample modes efficiently
along the $\mu_\mathbf{k}$ direction. 
Note that this effect cannot be mitigated even when we increase the number of independent random realizations,
since we sample exactly the same wavevectors from all the simulations unless we change the box size for different realizations.
Higher multipole moments ($\ell = 2$ and $4$ in our case) are more sensitive to this effect since the kernel 
$\mathcal{P}_\ell(\mu_\mathbf{k})$ strongly depends on the argument, $\mu_\mathbf{k}$.
This effect is discussed, e.g., in \citet{Nishimichi11,Taruya13} and its impact on the hexadecapole
is shown to be significant even when the simulation box size is as large as $\sim2\,h^{-1}\,\mathrm{Gpc}$.

\begin{figure}
\begin{center}
\includegraphics[height=7.4cm]{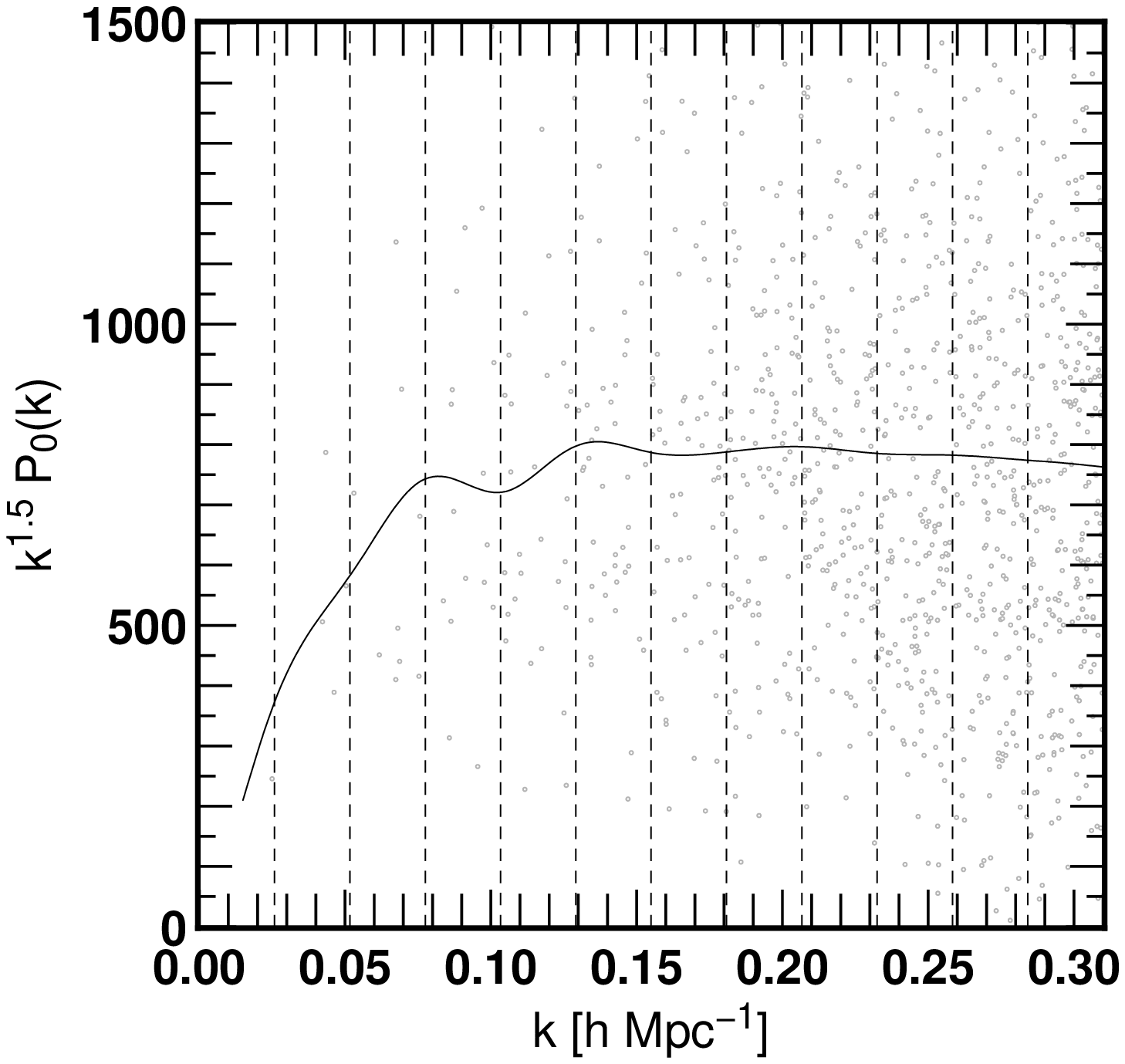}
\includegraphics[height=7.4cm]{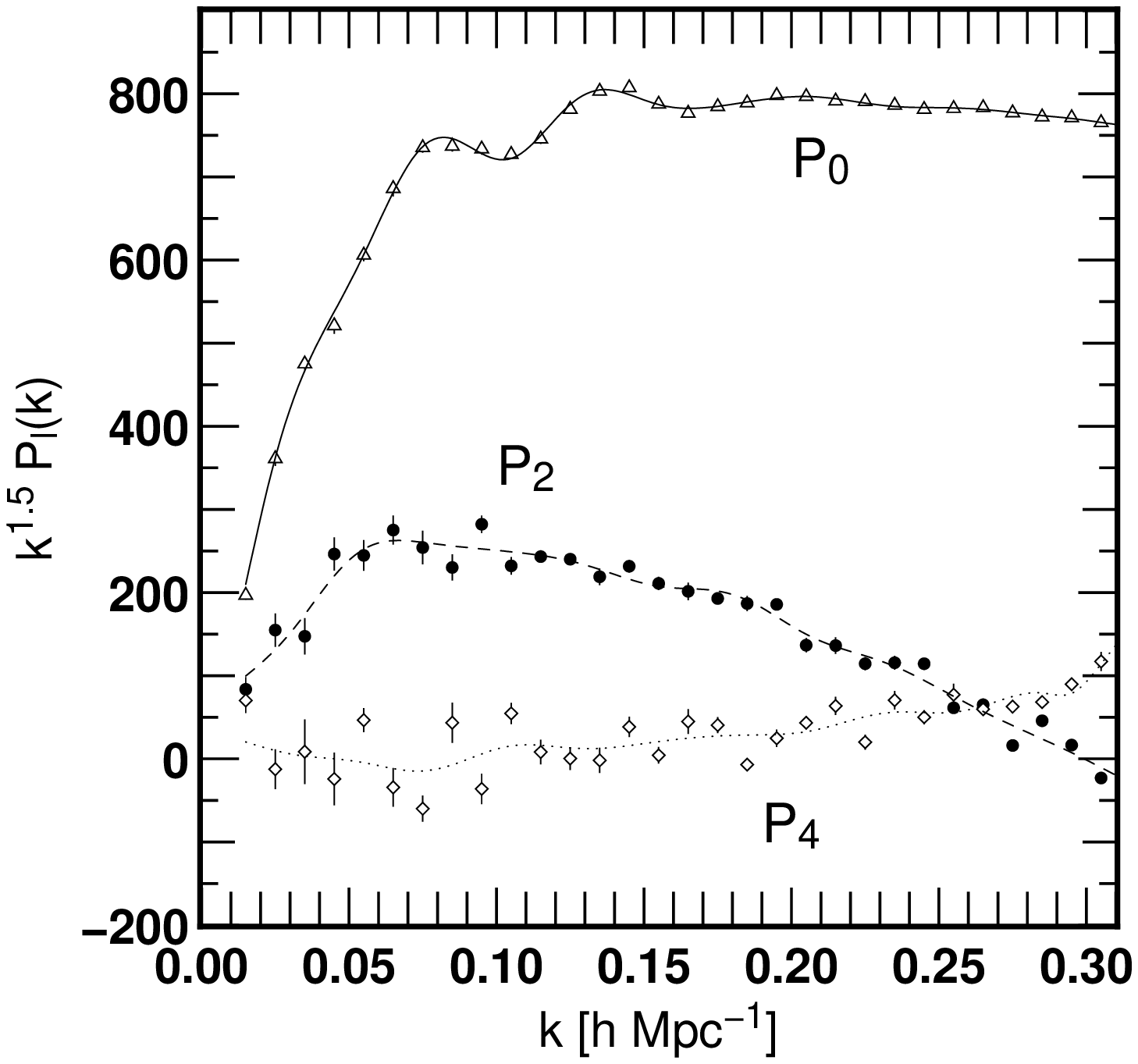}
\caption{Measurement of the power spectrum based on the cubic B-spline fitting.
We plot in the top panel a random sample of $1,000$ data points, $|\delta_\mathbf{k}^{(n)}(k)|$, out of $3,880,650$ fitted 
to compute the monopole moment (points) and the result of the fit (solid line). 
The vertical dashed lines show the positions of breakpoints used to construct the basis function.
The bottom panel shows a comparison of multipole moments of the power spectrum calculated from different estimators.
Lines show the multipoles based on the cubic B-spline fitting while the results of the binned estimator are depicted by
symbols with error bars.}
\label{fig:Bspline}
\end{center}
\end{figure}

In this study, we thus implement a different estimator of the multipole moments.
As outlined in Section~\ref{subsec:fit}, we apply the B-spline fitting to the {\it unbinned} weighted square of the density
contrast (i.e., equation~\ref{eq:unbinned}). This is demonstrated in the top panel of Fig.~\ref{fig:Bspline}. 
We explain the detail of this procedure here.
We choose $13$ breakpoints equally-spaced from $0$ to $0.31\,h\,\mathrm{Mpc}^{-1}$ to construct the basis function 
(vertical dashed lines).
This is chosen so that the resultant curve is smooth enough but the wiggles of BAOs are not degraded significantly.
We employ a cubic B-spline function and fit the data points shown by points.

With this procedure, the aforementioned artificial noise in the multipoles is greatly reduced.
We show the model power spectrum based on the B-spline fitting described above and
compare it with the binned power spectrum adopting the best-fit parameters of Model 5a.
The error bars on the binned power spectrum indicate the $1\sigma$ statistical error of the data points
expected from the scatter among different realizations. Higher multipole moments, especially the hexadecapole
moment, show a notable noisy pattern larger than the typical size of error bars.
On the other hand, the multipoles based on the B-spline fitting are much smoother.
We can still find clearly by eye the baryon acoustic signature on the monopole moment even after the B-spline fitting.

We next discuss the statistical accuracy of the model power spectrum.
One of an approximate treatment in this study is that we take account of
the statistical error only on the observed power spectrum but neglect that
on the model power spectrum estimated from finite simulation volume.
Even though the total volume of $11$ realizations ($16\,h^{-3}\,\mathrm{Gpc}^3$, and can be slightly different when
we consider the AP distortion) 
is larger than the observed volume ($1.39\,h^{-3}\,\mathrm{Gpc}^3$ in our fiducial cosmological model),
the accuracy of the model power spectrum should carefully be checked.

\begin{figure}
\begin{center}
\includegraphics[height=7.4cm]{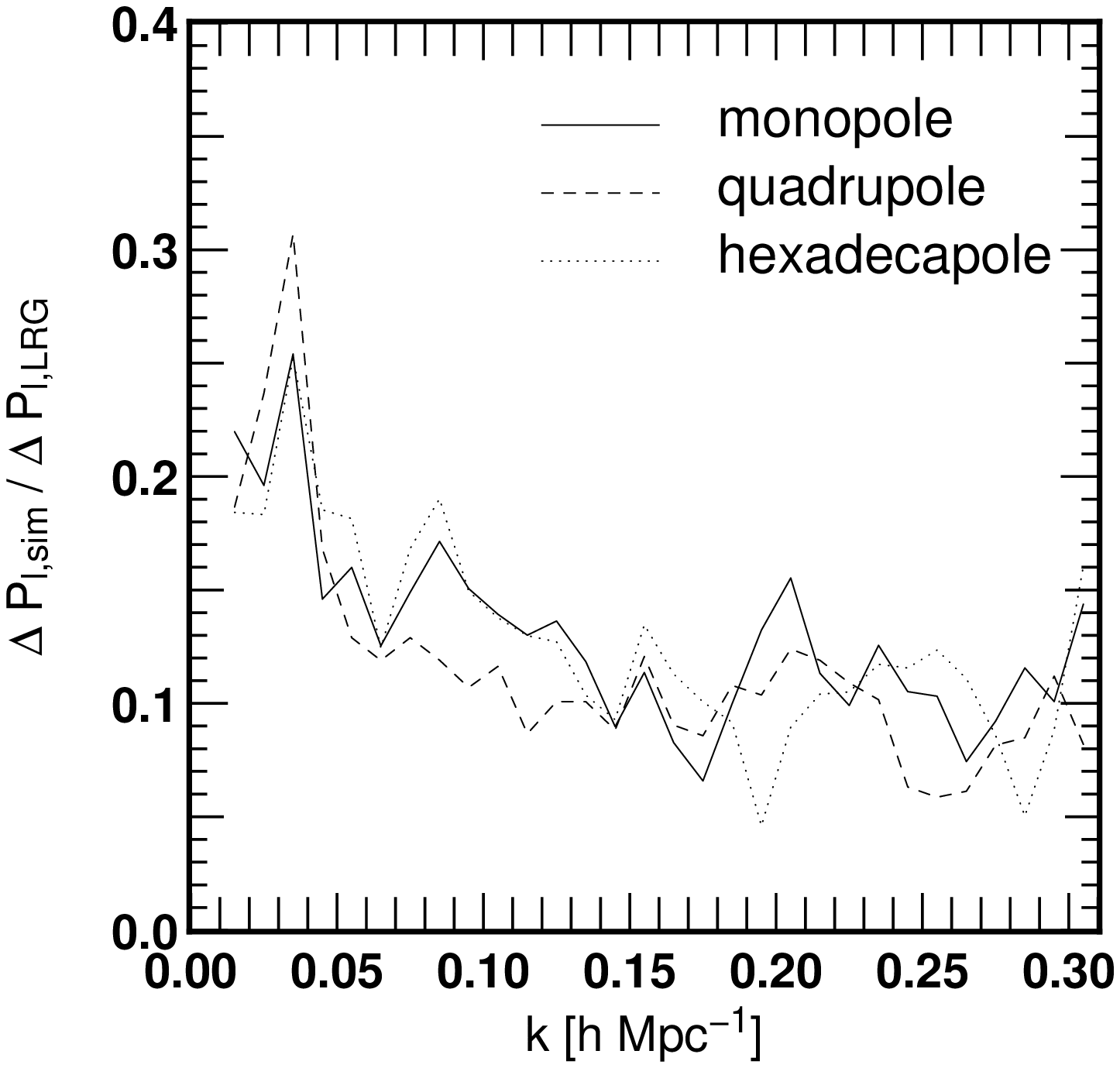}
\caption{Statistical error on the model power spectrum compared with the measured one.}
\label{fig:error}
\end{center}
\end{figure}

We plot in Fig.~\ref{fig:error} the statistical error on the simulated multipole moments of the
power spectrum divided by that on the observational data. In the plot, we again adopt the best-fit parameters
of Model 5a and estimate the statistical error on the model power spectrum by
\be
[\Delta{P}_{\ell,\mathrm{sim}}(k)]^2 &=& \frac{1}{N_\mathrm{r}-1}\sum_n \left[\hat{P}_{\ell}^{(n)}(k)-\bar{P}_{\ell}(k)\right]^2,
\label{eq:error}
\ee
where $N_\mathrm{r} = 11$ is the number of independent realizations and 
$\hat{P}_\ell^{(n)}$ and $\bar{P}_\ell(k)$ denote the $\ell$th multipole moment of the power spectrum 
measured from the $n$th realization and that averaged over $N_\mathrm{r}$ realizations, respectively.
We plot in Fig.~\ref{fig:error} $\Delta P_{\ell,\mathrm{sim}}(k)$ divided by the statistical error on the observed multipoles.

Overall, the ratio is $10$ to $30$ per cent depending on the wavenumber and is almost independent of multipoles.
We can roughly estimate it from the fact that the error bar scale as $\mathrm{volume}^{-1/2}$, and this gives
a $29$ per cent smaller error on the model multipoles. This is consistent with the result of Fig.~\ref{fig:error} on large scales
($k\sim0.03\,h\,\mathrm{Mpc}^{-1}$). On smaller scales since our mock LRGs have a larger number density than
the observed LRGs, the mock multipoles are less affected by shot noise. That is why the ratio decreases with wavenumber.

On the other hand, the estimate of the covariance matrix of the observed multipoles adopted in this study
is expected to be similarly accurate.
For example, the cross covariance between the different multipoles, which we ignore in this study, 
can be evaluated by linear theory, and the correlation coefficient for the monopole-quadrupole cross covariance is 
$\sim20$ per cent for $\beta=f/b\sim0.28$ (see Figure~4 of \citealt{Taruya11}).
Also, the non-Gaussian component of the covariance matrix is $\simlt30$ per cent in redshift space 
on the scale of our interest according to \citet{Takahashi09} in which they perform $5,000$ 
cosmological $N$-body simulation to estimate the covariance matrix (see Figures~2 and 3 in the reference).
We thus expect that the final error ellipses of the model parameters is accurate at this level ($10$ to $30$ per cent).

\section{Impact of fiber collisions}
\label{sec:fiber}
We discuss the possible impact of the fiber collisions on the multipole moments of the power spectrum in this Appendix.
The LRG sample analyzed in \citet{Yamamoto10} suffers from spectroscopic fiber collisions, 
and some targeted LRG candidates do not have spectra when they have a close companion within $55\,\mathrm{arcsec}$.
Although both of close pairs have redshifts thanks to multiple observations in some region of the sky,
the redshifts about $7$ per cent of the targeted objects are not measured.
In the measurement of the power spectrum, \citet{Yamamoto10} do not make a correction for fiber collisions for simplicity.
However, since fiber collisions occur pairs of galaxies that are very close in the plane-of-sky direction, they can
be a source of apparent anisotropy of the clustering of galaxies. This effect is studied in \citet{Guo12} for
the multipole moment of the two-point correlation function. Although they find that it is significant only
on small scales ($\simlt 10\,h^{-1}\mathrm{Mpc}$), its impact on the Fourier counterpart is not trivial.
Here we give a rough estimate of this effect using the mock LRGs.

We use mock LRGs in our simulation and simulate this effect by searching for fiber collision paris and randomly
discarding one of two galaxies subject to a fiber collision. There are small number of mock LRGs that have
more than one companions that are closer than $55\,\mathrm{arcmin}$, and we discard these mock LRGs
preferentially to reduce the total number of discarded mock LRGs.
We keep the distant observer approximation and convert the separation of pairs perpendicular to the line of sight
to the observed angle assuming the angular diameter distance at the effective redshift of observation, $z=0.3$.
We do not take into account of multiple observations of plates to assess the maximum impact of fiber collisions.
Strictly speaking, we should redo the analysis by including this effect into the MCMC analysis to be more consistent.
However, we simply take the best-fit mock catalog explained in the main text (Model 5a) and then implement
the effect of fiber collisions since a MCMC analysis with this effect is time consuming.

\begin{figure}
\begin{center}
\includegraphics[height=7.4cm]{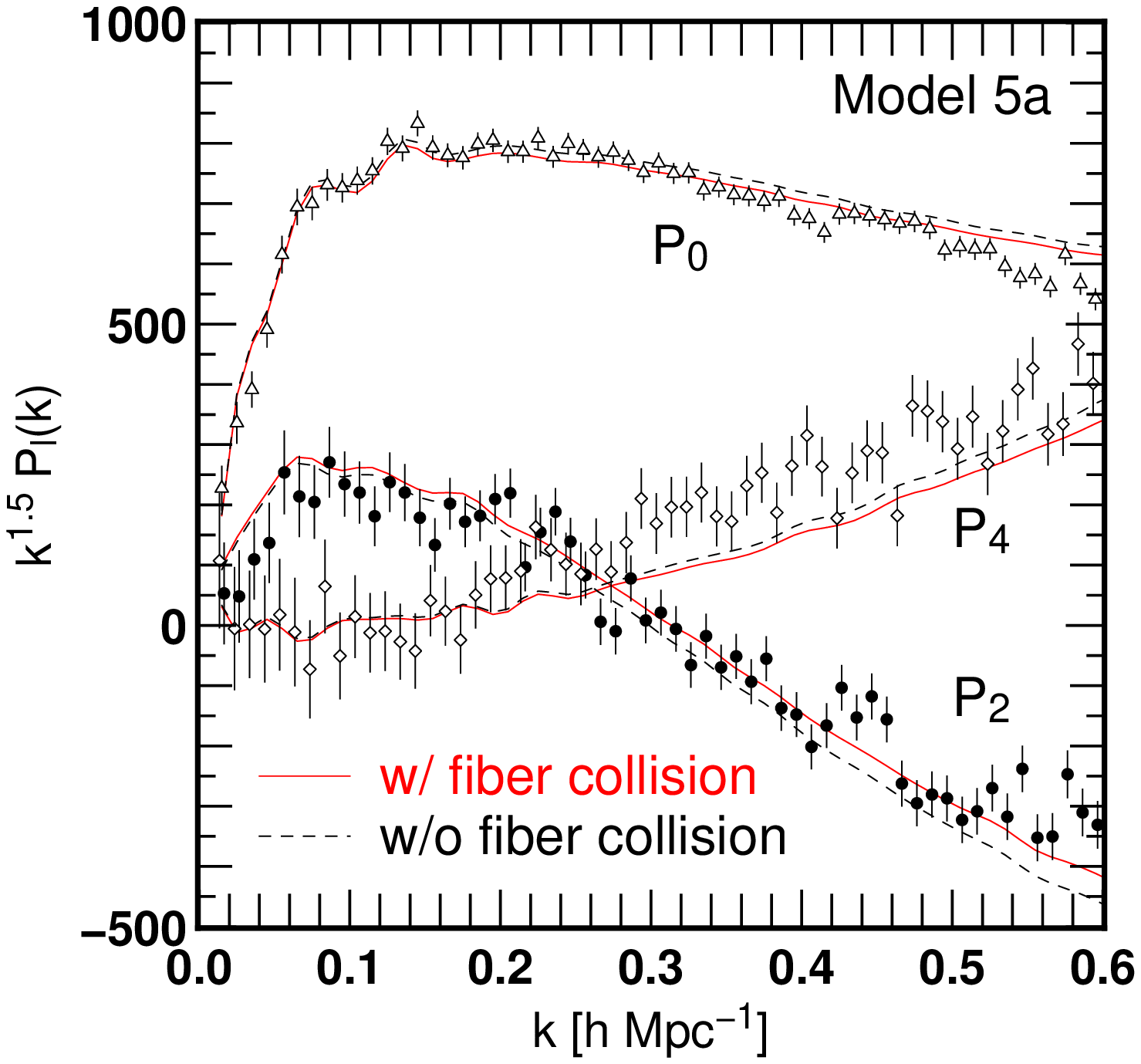}
\caption{Impact of fiber collisions on the power spectrum in redshift space.
We plot the multipole moments measured from our best-fit mock LRGs (Model 5a)
with (solid) and without (dashed) fiber collisions.}
\label{fig:fb}
\end{center}
\end{figure}

We show the result in Fig.~\ref{fig:fb}. Notice that we plot a wider wavenumber range than in the main text 
to see the impact of fiber collisions down to smaller scales, but the fitting range is same as before 
(i.e., $k_\mathrm{max}=0.305\,h\,\mathrm{Mpc}^{-1}$). 
The figure shows the multipole moments with
or without fiber collisions (solid and dashed lines, respectively). We also plot the observational data by symbols with
error bars. Fiber collisions suppress the monopole and hexadecapole, and amplify the quadrupole moment.
The effect is more prominent on smaller scales.
However, at $k\simlt0.3\,h\,\mathrm{Mpc}^{-1}$, the difference between the solid and the dashed lines are typically
much smaller than the statistical error on the observational data.
Thus we expect that fiber collisions do not affect the parameter constraints significantly as long as we focus on
relatively large scales. Since our fitting procedure using $N$-body data is in principle applicable to nonlinear scales unlike
perturbation theory based models that break down on certain scales where nonlinearity is very strong, 
a proper inclusion of fiber collisions in the analysis might be important to enlarge the wavenumber range 
from which we can extract cosmological information robustly.

\end{document}